\documentclass[useAMS,usenatbib]{mn2e}

\usepackage[a4paper,margin=2cm]{geometry}
\usepackage{graphicx}
\usepackage{lscape}
\usepackage{amssymb,amsmath,mathrsfs}
\usepackage{subfig}
\usepackage{color}
\usepackage{threeparttable}

\renewcommand{\tabcolsep}{4.5pt}

\def\aj{AJ}                   
\def\apj{ApJ}                 
\def\apjl{ApJ}                
\def\aap{A\&A}                
\def\aapr{A\&ARv.}          
\def\mnras{MNRAS}             
\def\pasa{Publ. Astron. Soc. Australia}  
\def\iauc{IAU Circ.}       
\def\aspcs{ASP Conf. Ser.}       
\def\ea{Exp. Astron.}       
\def\assp{Ap\&SS Proc.}       

\title[H\,{\normalsize \it I} in nearby galaxies]
  {H\,{\Large\bf I} emission and absorption in nearby, gas-rich galaxies}
\author[S. N. Reeves et al.]
  {S. N.~Reeves,$^1$$^,$$^2$$^,$$^3$\thanks{E-mail: sarah@physics.usyd.edu.au}
  E. M.~Sadler,$^1$$^,$$^3$ J. R.~Allison,$^1$$^,$$^2$$^,$$^3$ B. S.~Koribalski,$^2$
   \newauthor 
  S. J.~Curran,$^1$$^,$$^3$$^,$$^4$ and M. B.~Pracy$^1$\\
  $^1$Sydney Institute for Astronomy, School of Physics A28, The University of Sydney, NSW 2006, Australia\\
  $^2$Australia Telescope National Facility, CSIRO Astronomy and Space Science, PO Box 76, Epping, NSW 1710, Australia\\
   $^3$ARC Centre of Excellence for All-Sky Astrophysics (CAASTRO)\\
   $^4$School of Chemical and Physical Sciences, Victoria University of Wellington, PO Box 600, Wellington 6140, New Zealand }
\date{Released 2015 Xxxxx XX}

\pagerange{\pageref{firstpage}--\pageref{lastpage}} \pubyear{2015}

\def\LaTeX{L\kern-.36em\raise.3ex\hbox{a}\kern-.15em
    T\kern-.1667em\lower.7ex\hbox{E}\kern-.125emX}

\begin{document}

\label{firstpage}

\maketitle

\begin{abstract}
We present the results of a targeted search for intervening \mbox{H\,{\sc i}} absorption in six nearby, gas-rich galaxies using the Australia Telescope Compact Array (ATCA). 
The sightlines searched have impact parameters of 10-20 kpc. 
By targeting nearby galaxies we are also able to map their \mbox{H\,{\sc i}} emission, allowing us to directly relate the absorption-line detection rate to the extended \mbox{H\,{\sc i}} distribution. 
The continuum sightlines intersect the \mbox{H\,{\sc i}} disk in four of the six galaxies, but no intervening absorption was detected. 
Of these four galaxies, we find that three of the non-detections are the result of the background source being too faint. 
In the fourth case we find that the ratio of the spin temperature to the covering factor ($T_{\mathrm{S}}/f$) must be much higher than expected ($\gtrsim$5700 K) in order to explain the non-detection. 
We discuss how the structure of the background continuum sources may have affected the detection rate of \mbox{H\,{\sc i}} absorption in our sample, and the possible implications for future surveys. 
Future work including an expanded sample, and VLBI observations, would allow us to better investigate the expected detection rate, and influence of background source structure, on the results of future surveys.
\end{abstract}


\section{Introduction}
\label{introduction}

Observations of the 21 cm transition of neutral atomic hydrogen (\mbox{H\,{\sc i}}) with radio telescopes provide a unique probe of the cool gas in and around galaxies. 
Most spirals possess an \mbox{H\,{\sc i}} disk, typically around twice the size of the stellar component seen in visible light \citep{1981AJ.....86.1825B}. 
The 21 cm line can trace large-scale tidal features --- tails and bridges --- which are evidence of past or on-going interactions between galaxies (see e.g. \citealt{2001ASPC..240..657H}).
\mbox{H\,{\sc i}} observations therefore provide crucial information about the size, mass and structure of galaxies at different epochs, and are an essential tool for understanding how galaxies have evolved over cosmic time.

Observations of nearby galaxies allow us to observe these features directly, by their 21 cm emission-line. 
Recent surveys such as the Local Volume \mbox{H\,{\sc i}} Survey \citep[LVHIS,][]{2008glv..book...41K} and The \mbox{H\,{\sc i}} Nearby Galaxy Survey \citep[THINGS,][]{2008AJ....136.2563W} have used radio interferometers to map the \mbox{H\,{\sc i}} gas with high spatial and spectral resolution, allowing us to study the distribution and kinematics of the gas in unprecedented detail. 
However, such observations are time consuming, and can only be conducted for a relatively small number of galaxies. 

At the other extreme, large surveys with single dish telescopes provide a much broader picture of the local galaxy population, though without the resolution required for studying individual galaxies in detail. 
The \mbox{H\,{\sc i}} Parkes All Sky Survey \citep[HIPASS,][]{2001MNRAS.322..486B,2004AJ....128...16K,2004MNRAS.350.1195M,2006MNRAS.371.1855W} detected over 5000 galaxies at redshifts $z < 0.04$ resulting in the most robust calculation of the \mbox{H\,{\sc i}} Mass Function (HIMF) at that time \citep{2005MNRAS.359L..30Z}. 
The Arecibo Legacy Fast ALFA Survey \cite[ALFALFA,][]{2005AJ....130.2598G,2011AJ....142..170H} currently being conducted with the 305m Arecibo telescope will extend this to higher redshifts, and has already detected 15,000 galaxies at redshifts $z < 0.06$.

However, the detectability of the 21 cm emission-line drops off rapidly with redshift and to date the most distant detection of a galaxy in emission is $z = 0.25$ \citep{2008ApJ...685L..13C}. 
\mbox{H\,{\sc i}} stacking experiments, using optical redshift data, allow us to make statistical detections in existing data which are not deep enough to detect individual galaxies (see e.g. \citealt{2013MNRAS.433.1398D}).
This provides useful information about the average properties of galaxies but it is impossible to learn anything about the individual galaxies that have been stacked, and making detections at higher redshifts is still extremely difficult.

The 21 cm line can instead be detected in absorption, seen as a negative feature in the spectrum of a background radio continuum source. 
The detectability of the \mbox{H\,{\sc i}} absorption-line is essentially independent of distance, provided there is a sufficiently bright background source against which to search. 
This makes it possible to study neutral gas at much higher redshifts than is possible in emission. 

Intervening absorption occurs when a reservoir of gas is found along the sightline to an unassociated background continuum source. 
Detections of intervening absorption are therefore relatively rare, since they require the chance alignment of the foreground gas cloud with a sufficiently bright background source. 
As a result, there have been limited attempts to conduct blind searches for intervening absorption, with most searches focusing on targeted observations of bright radio sources where the sightline is known to pass close to a foreground galaxy.

The largest blind \mbox{H\,{\sc i}} absorption-line survey to date covers an area of 517 deg$^{2}$ at redshifts $z \lesssim 0.06$ \citep{2011ApJ...742...60D}. 
Making use of the first ALFALFA data, sightlines to 8983 radio sources were searched for absorption. 
They re-detected one associated absorption-line, but did not detect any intervening absorption-lines. 
Upper limits on the \mbox{H\,{\sc i}} column density were found to be consistent with previous estimates of the HIMF, and they conclude that higher sensitivity surveys will likely result in numerous \mbox{H\,{\sc i}} absorption-line detections.

Next-generation radio telescopes, with their unprecedented bandwidth, field-of-view, and sensitivity will make it possible to conduct the first large-scale blind absorption-line surveys. 
FLASH\footnote{http://www.physics.usyd.edu.au/sifa/Main/FLASH/} (the `First Large Absorption Survey in \mbox{H\,{\sc i}}') is one such survey. 
FLASH will use the Australian Square Kilometre Array Pathfinder \citep[ASKAP,][]{2008ExA....22..151J} to search for \mbox{H\,{\sc i}} absorption along 150,000 sightlines, studying the evolution of neutral hydrogen in the redshift range $0.5 < z < 1.0$ (look-back times of $\sim$4-8 Gyr).

However, in order to relate absorption-line data to physical galaxy properties, at redshifts where emission-line data will not be available, we need to know how the detection rate of absorption varies with linear distance from the centre of the galaxy (i.e. impact parameter). 
Two recent surveys have attempted to address this question.

\citet{2010MNRAS.408..849G} searched for intervening absorption in five quasar galaxy pairs (QGPs) with impact parameters of 11.0-53.0 kpc, and detected one absorption-line at an impact parameter of 11.0 kpc. 
Combining their results with those in the literature --- 15 additional quasar sightlines from six different surveys \citep{1975ApJ...200L.137H,1988A&A...191..193B,1990ApJ...356...14C,1992ApJ...399..373C,2004ApJ...600...52H,2010ApJ...713..131B} ---  they estimate that there is a 50 per cent chance of detecting an absorption-line within 20 kpc of a galaxy, for integrated optical depths $\tau > 0.1$ km s$^{-1}$.

A complementary study searched for \mbox{H\,{\sc i}} absorption in fifteen QGPs at impact parameters of 1.7-86.7 kpc, making  two new detections, at 1.7 and 2.8 kpc \citep{2010ApJ...713..131B,2011ApJ...727...52B,2014ApJ...795...98B}. 
In agreement with the \citet{2010MNRAS.408..849G} results, it also appears that detections at impact parameters greater than 20 kpc are very rare.

Here, we expand on previous work by conducting a targeted search for intervening absorption in a sample of nearby, gas-rich galaxies. 
By selecting low-redshift galaxies we are also able to map the target galaxies in \mbox{H\,{\sc i}}, allowing us to directly relate the gas distribution to the absorption-line detection rate. 
The combined emission- and absorption-line data also allows us to put some constraints on quantities such as the spin temperature and covering factor of the gas, which cannot be measured from absorption-line data alone.

Throughout this work we assume a flat $\Lambda$CDM cosmology, with $\Omega_{\mathrm{M}}$ = 0.27, $\Omega_{\Lambda}$ = 0.73, and $H_{\mathrm{0}}$ = 71 km s$^{-1}$ Mpc$^{-1}$. 
All uncertainties refer to the 68.3 per cent confidence interval, unless
otherwise stated. 
We use the optical definition of the velocity ($v = cz$) throughout.

\section{Observations and data reduction}
\label{data}

\subsection{Sample selection}
\label{data:sample_selection}

\begin{table*}
\begin{minipage}{\linewidth}
\centering
\caption{\mbox{H\,{\sc i}} and optical properties of the target galaxies. 
Columns (1)-(6) are the \mbox{H\,{\sc i}} properties as given in the HIPASS BGC \citep{2004AJ....128...16K}. 
Column (1) is the HIPASS name. 
Columns (2) and (3) are the HIPASS J2000 right ascension and declination. 
Columns (4) and (5) are the systemic and local group velocities. 
Column (6) is the optical/IR identification. 
Columns (7)-(9) are the optical properties as given in the NASA Extragalatic Database (NED). 
Column (7) is the optical morphological classification. 
Columns (8) and (9) are the optical J2000 right ascension and declination.}
\label{table:hipass}
\begin{threeparttable}
\begin{tabular}{@{} lllrrlllll @{}} 
\hline
\multicolumn{6}{l}{HIPASS} & \multicolumn{3}{l}{Optical} \\
\hline
HIPASS Name & RA & Dec. & $v_{\mathrm{sys}}$ & $v_{\mathrm{LG}}$ & Optical ID & Morph. Type & RA & Dec. \\
& (J2000) & (J2000) & (km s$^{-1}$) & (km s$^{-1}$)& & & (J2000) & (J2000) \\
\hline
J0022-53 & 00 22 30 & $-$53 38 55 & 1436 & 1335 & ESO\,150-G\,005 & SAB(s)dm & 00 22 25.56 & $-$53 38 51.0\tnote{$1$} \\
J2243-39 & 22 43 09 & $-$39 51 01 & 2149 & 2147 & ESO\,345-G\,046 & SA(s)d & 22 43 16.09 & $-$39 51 58.9\tnote{$1$} \\
J2122-36 & 21 22 18 & $-$36 37 50 & 2573 & 2599 & ESO\,402-G\,025 & Sp. (unclass.) & 21 22 15.79 & $-$36 37 39.6\tnote{$2$} \\
J0331-51 & 03 31 41 & $-$51 54 20 & 1064 & 877 & IC\,1954 & SB(s)b & 03 31 31.39 & $-$51 54 17.4\tnote{$2$} \\
J2255-42 & 22 55 47 & $-$42 38 30 & 1710 & 1690 & NGC\,7412 & SB(s)b &  22 55 45.75 & $-$42 38 31.3\tnote{$2$} \\
J2257-41 & 22 57 18 & $-$41 04 02 & 939 & 927 & NGC\,7424 & SAB(rs)cd & 22 57 18.37	& $-$41 04 14.1\tnote{$2$} \\
\hline
\end{tabular}
\begin{tablenotes}
\footnotesize{
\item[] {Optical position references:}
\item[$1$] {\citet{1996MNRAS.278.1025L}}
\item[$2$] {\citet{2006AJ....131.1163S}}
}
\end{tablenotes}
\end{threeparttable}
\end{minipage}
\end{table*}

\begin{table*}
\begin{minipage}{\linewidth}
\centering
\caption{Properties of the background continuum sources used to search for absorption. 
Column (1) is the identification (ID) used throughout this work. 
Columns (2)-(6) are the radio properties as given in the SUMSS catalogue \citep{2003MNRAS.342.1117M}. 
Column (2) is the SUMSS name. 
Columns (3) and (4) are the SUMSS J2000 right ascension and declination.
Columns (5) and (6) are the 843 MHz peak and integrated fluxes. 
Columns (7) and (8) are the angular and linear separations of the radio source and the foreground galaxy.}
\label{table:sumss}
\begin{threeparttable}
\begin{tabular}{@{} llllrrrr@{}} 
\hline
Source Name\tnote{$*$} & SUMSS Name & RA & Dec & $S_{\mathrm{peak,843}}$ & $S_{\mathrm{int,843}}$ & Ang. Sep. & Lin. Sep. \\
& & (J2000) & (J2000) & (mJy bm$^{-1}$) & (mJy) & (arcmin) & (kpc) \\
\hline
C-ESO\,150-G\,005-1 & J002243-533644 & 00 22 43.80 & $-$53 36 44.00 & 96.5 & 126.0 & 3.4 & 18.7 \\
C-ESO\,150-G\,005-2 & J002248-533548 & 00 22 48.05 & $-$53 35 49.0 & 69.3 & 100.6 & 4.5 & 24.5 \\
C-ESO\,345-G\,046 & J224313-395029	 & 22 43 13.82 & $-$39 50 29.90 & 76.2 & 77.5 & 1.5 & 13.5 \\
C-ESO\,402-G\,025 & J212223-363806 & 21 22 23.62 & $-$36 38 6.60 & 162.4 & 219.5 & 1.6 & 17.2 \\
C-IC\,1954 & J033145-515618 & 03 31 45.66 & $-$51 56 18.00 & 257.6 & 278.2 & 3.0 & 10.7 \\
C-NGC\,7412 & J225536-423746 & 22 55 36.32 & $-$42 37 46.50 & 176.9 & 203.4 & 1.9 & 13.0 \\
C-NGC\,7424 & J225729-410241 & 22 57 29.56 & $-$41 02 41.20 & 78.7 & 82.6 & 2.6 & 9.9 \\
\hline
\end{tabular}
\begin{tablenotes}
\footnotesize{
\item[$*$] {Throughout this work we refer to the continuum sources by the ID listed in Column (1). This is the galaxy name, with a 'C' (for `continuum') prepended, allowing us to easily identify the galaxy-continuum source pairs. 
Where the background source consists of multiple components we have added a suffix, numbering the components in order of brightness.}
}
\end{tablenotes}
\end{threeparttable}
\end{minipage}
\end{table*}

\begin{figure*}
\hspace{-0.30cm}
\includegraphics[width=0.30\linewidth]{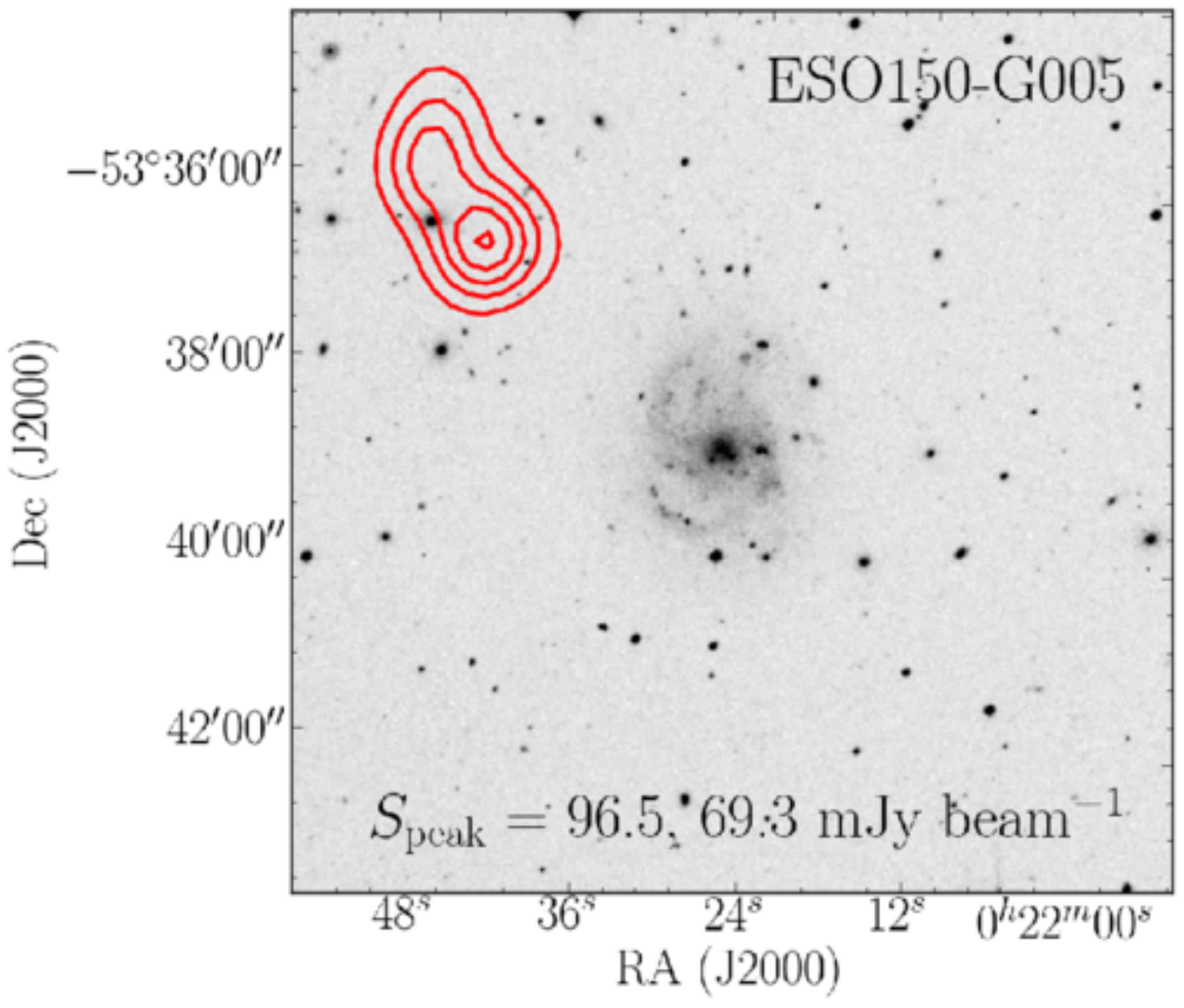}
\includegraphics[width=0.30\linewidth]{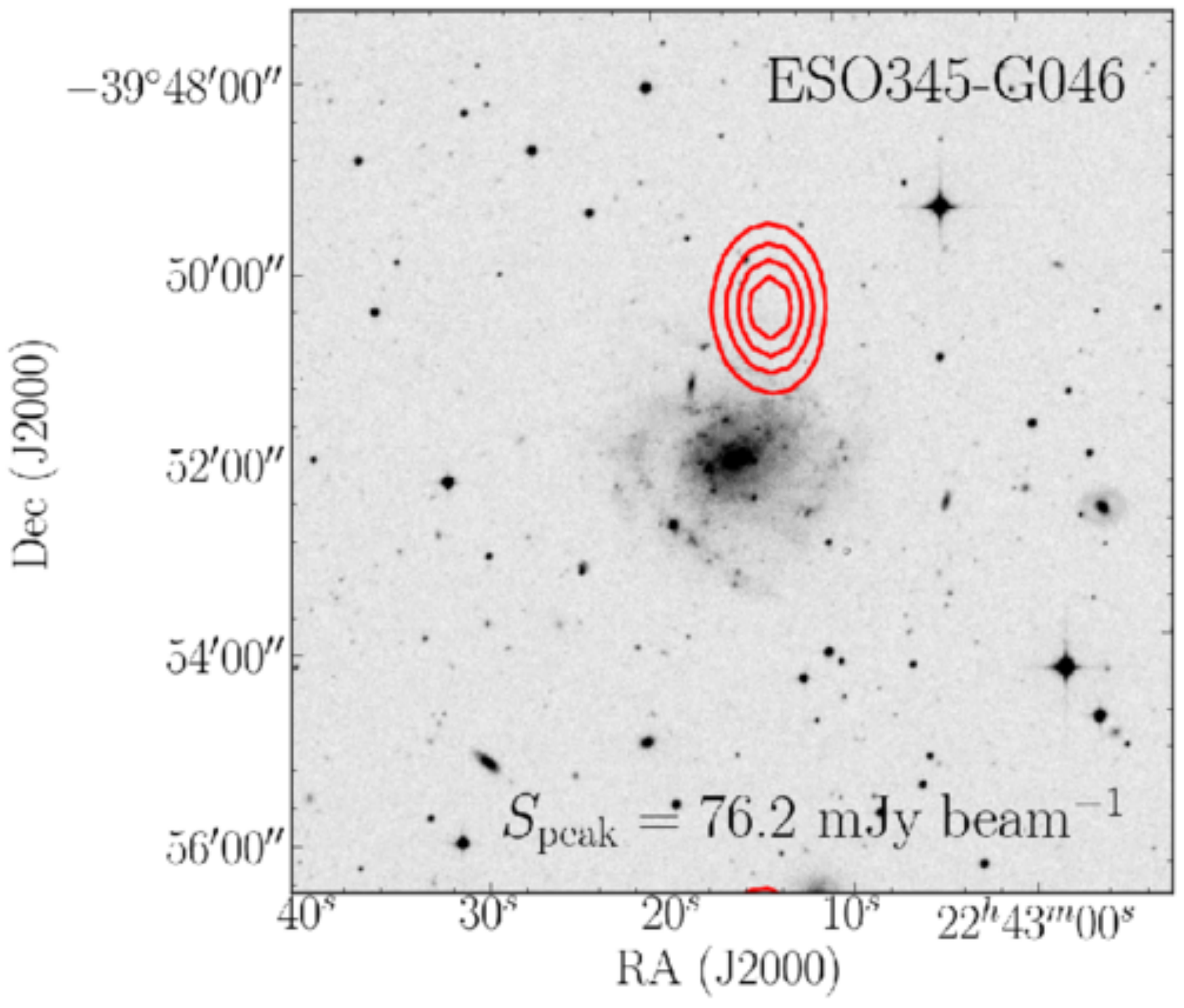}
\includegraphics[width=0.30\linewidth]{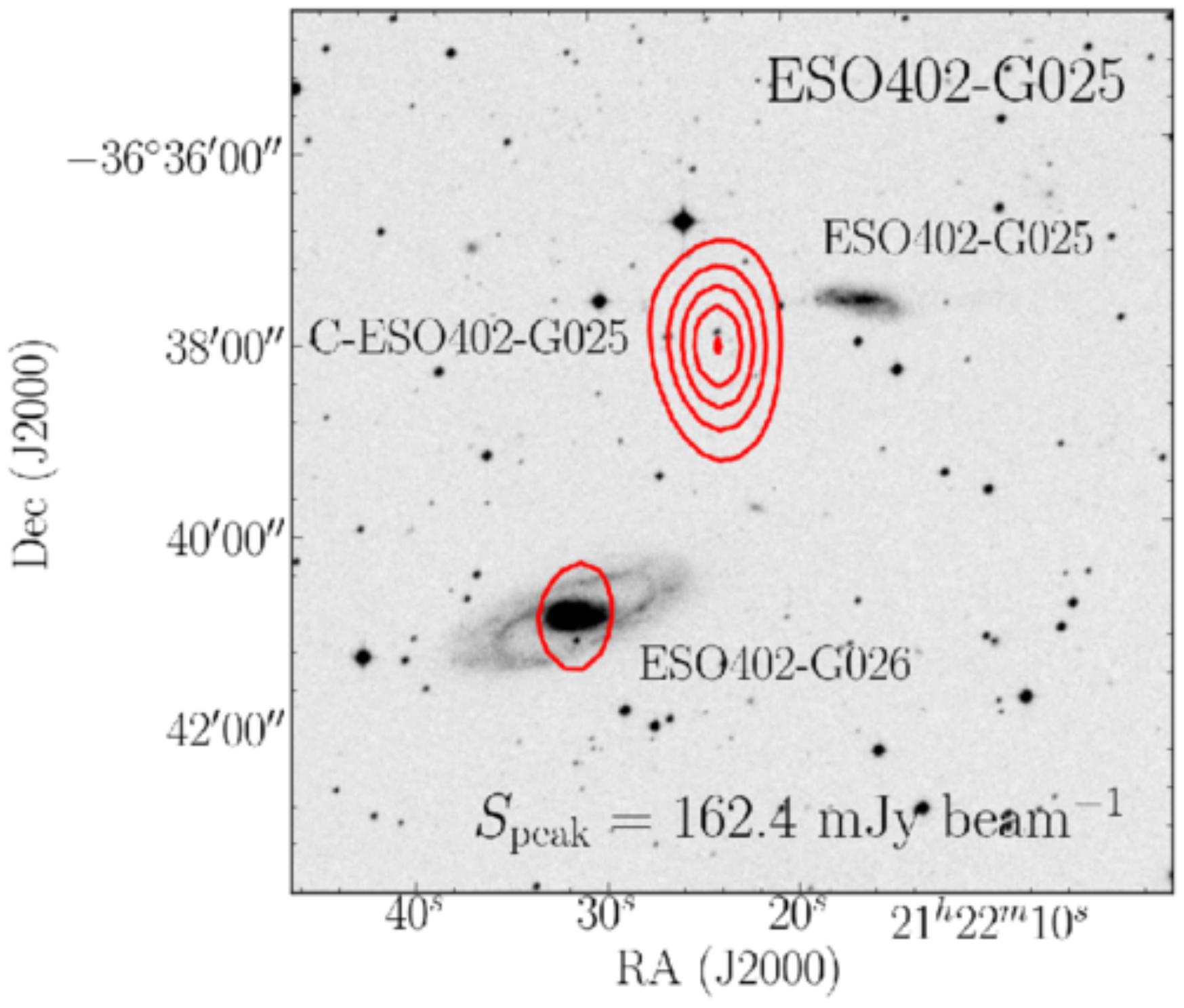}
\includegraphics[width=0.30\linewidth]{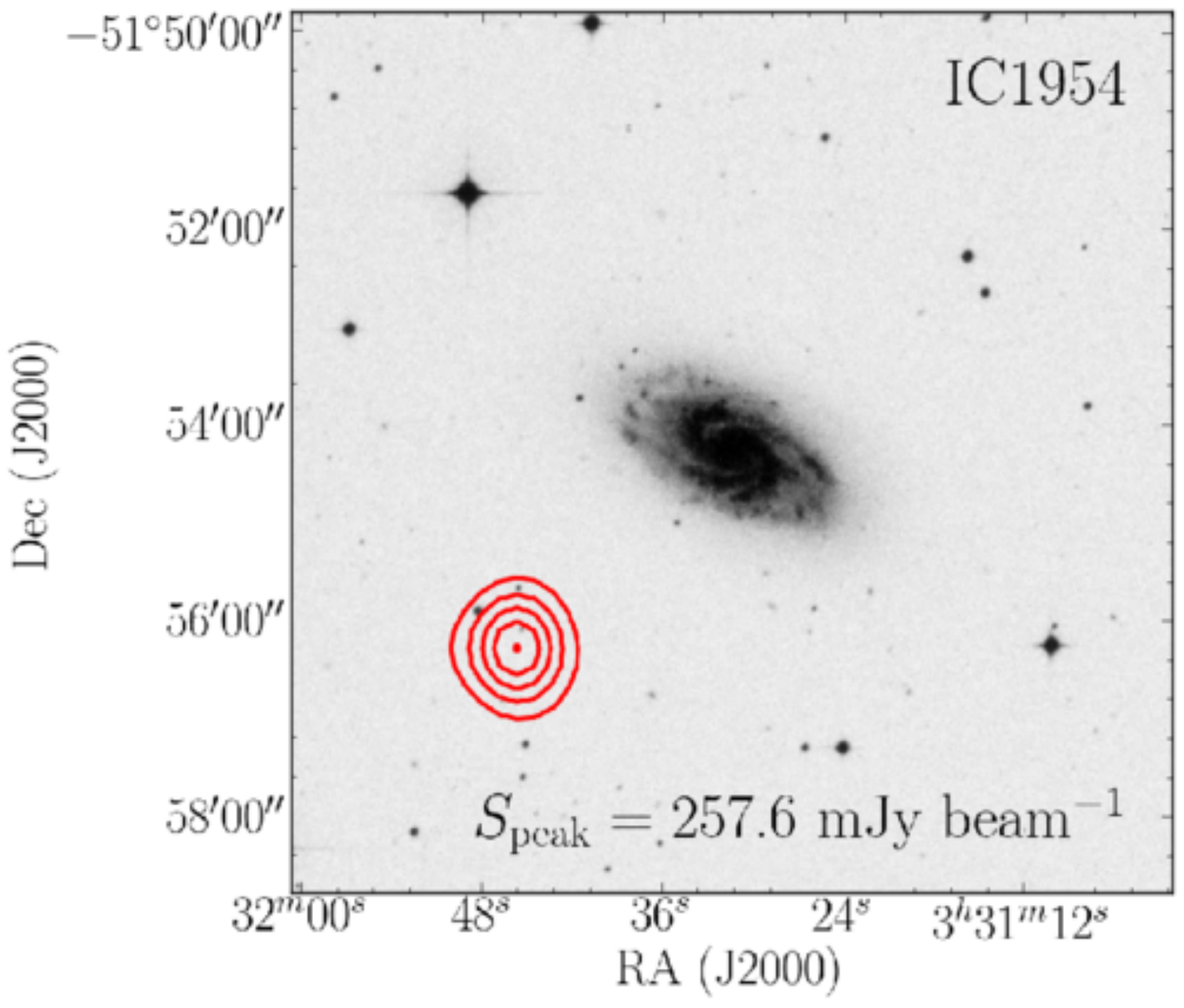}
\includegraphics[width=0.30\linewidth]{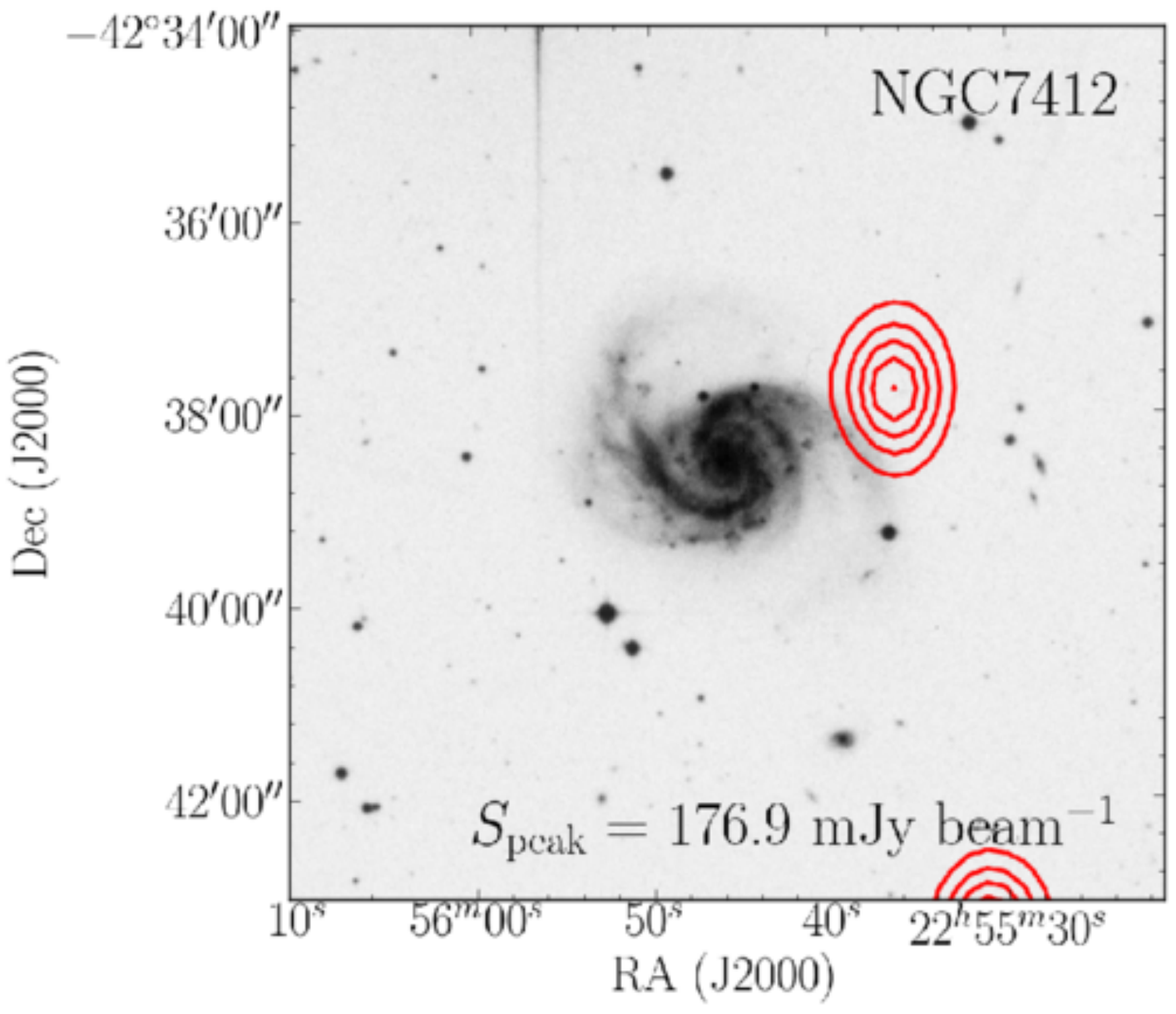}
\includegraphics[width=0.30\linewidth]{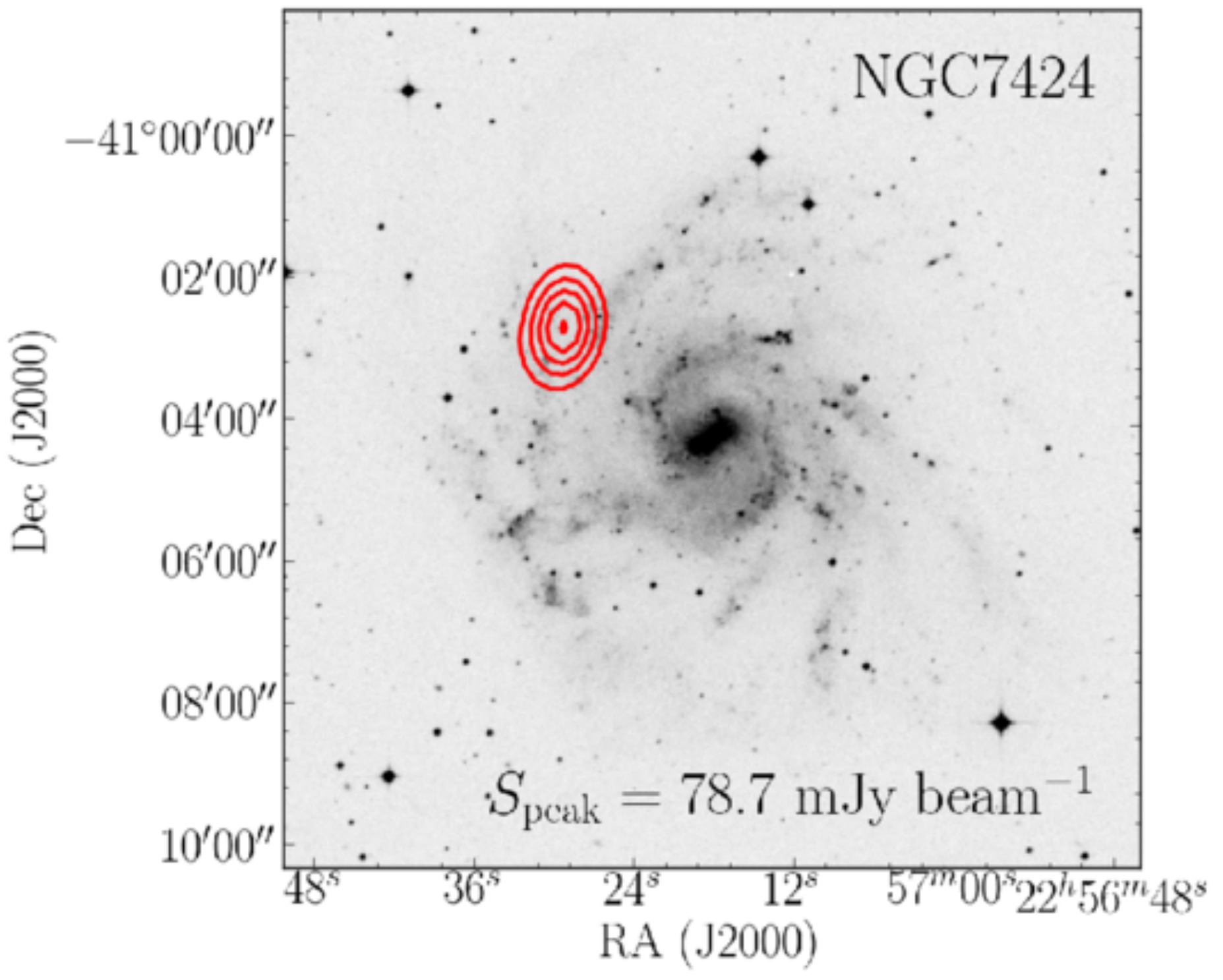}
\caption{SuperCOSMOS B-band images \citep{2001MNRAS.326.1279H} of the six target galaxies, with SUMSS 843 MHz radio continuum contours overlaid \citep{2003MNRAS.342.1117M}. 
The SUMSS peak flux is given in the bottom right corner. 
Radio contours start at 90 per cent of the peak flux and decrease in 10 per cent increments.} 
\label{figure:targets}
\end{figure*}

Our sample consists of six radio source-galaxy pairs with impact parameters of 20 kpc or less. 
These pairs were selected by cross-matching the HIPASS Bright Galaxy Catalogue \citep[HIPASS BGC,][]{2004AJ....128...16K} with the Sydney University Molonglo Sky Survey catalogue \citep[SUMSS,][]{2003MNRAS.342.1117M}. 
The BGC presents the 1000 \mbox{H\,{\sc i}}-brightest galaxies detected in the main HIPASS survey ($\delta < 0.0^{\circ}$), and so provides an ideal sample of nearby ($z < 0.04$), gas-rich galaxies from which to select our sample. 
Meanwhile, the SUMSS catalogue provides a radio continuum catalogue of the whole southern sky at 843 MHz from which to select background radio sources to search for absorption against. 

While we know some properties of the galaxies, such as their \mbox{H\,{\sc i}} masses, from HIPASS, little is known about the extent of the \mbox{H\,{\sc i}} disks. 
The 20 kpc impact parameter cutoff was chosen as (i) it is comparable to the typical size of \mbox{H\,{\sc i}} disks in bright spiral galaxies, and so identifies systems where the sightline is likely to pass through the \mbox{H\,{\sc i}} disk (good candidates for intervening absorption), and (ii) it targets the impact parameter range where previous searches indicate there is the highest chance of making a detection. 

Since the brightness of the background continuum source sets the detection limit for \mbox{H\,{\sc i}} absorption (for a given column density, spin temperature and covering factor), we have imposed a minimum flux density criterion of $S_{\mathrm{int,843}} > 50$ mJy (expected 1.4 GHz fluxes of $S_{\mathrm{int,1.4}} \gtrsim 35$ mJy, for a typical spectral index of $\alpha$ $\sim$ $-$0.7). 
The angular and linear separations were calculated using the optical galaxy positions. 
It is assumed that all of the continuum sources are genuine background sources, something we discuss further in Section \ref{discussion:detection_rate}.

In total the cross-matching process identified around 100 potential targets. 
The final selection of targets was based on a number of considerations --- we wished to select sources in a similar local sidereal time range (to aid in scheduling the observations) and also to prioritise those targets with the brightest background sources in order to maximise our detection rate. 

The final sample consists of six targets, with impact parameters in the range 10-20 kpc. 
The \mbox{H\,{\sc i}} properties of the target galaxies are presented in Table \ref{table:hipass}, and the properties of the background continuum sources in Table \ref{table:sumss}. 
The background source behind NGC\,7424 is a known quasar (and also an ultra-luminous X-ray source, \citealt{2004MNRAS.349.1093R,2006MNRAS.370.1666S}), but all of the other radio sources are unclassified. 
The field of ESO\,402-G\,025 contains a second nearby galaxy, ESO\,402-G\,026, which makes HIPASS J2122-36 a galaxy pair, with both galaxies potentially being close enough to detect intervening absorption.
Images of the six targets are shown in Figure \ref{figure:targets}.

\subsection{ATCA observations}
\label{data:observations}

\mbox{H\,{\sc i}} observations were carried out using the Australia Telescope Compact Array (ATCA) in the 750A and 750C array configurations (see Table \ref{table:atca_obs}). 
In these configurations, the movable antennas (CA01-CA05) have a maximum baseline of 750 m, with the fixed antenna (CA06) providing five longer baselines, which corresponds to an angular scale sensitivity of a few arcseconds to $\sim$10-20 arcmin). 
We used the Compact Array Broadband Backend \citep[CABB,][]{2011MNRAS.416..832W} zoom modes, in the 64M-32k configuration. 
With this setup we achieve 2048 channels over the 64 MHz `zoom-band', giving a spectral resolution of $\sim$6.7 km s$^{-1}$ at the centre of the band, where the \mbox{H\,{\sc i}} line is expected.

The observations were conducted in 6 $\times$ 12 hour periods between October 2011 and July 2012. 
The noise level achieved in a 12 hour observation ($\sim$2-3 mJy per channel) would be sufficient to detect \mbox{H\,{\sc i}} absorption in DLA-strength systems ($N_{\mathrm{HI}} \gtrsim 2 \times 10^{20}$) against the selected background radio sources ($S_{\mathrm{int,843}} > 50$ mJy), assuming they are unresolved.
The target sources were observed in 40 minute scans, interleaved with 10 minute scans of a nearby phase calibrator. 
The ATCA primary flux calibrator PKS1934-638 was observed for about 10-12 minutes at the beginning and end of each observation.

Observations were conducted at night to avoid the affects of solar interference, particularly on the short baselines. 
Some time was lost during the observations of NGC\,7412, due to poor weather. 
Due to a storm-related fault, antenna 6 was also offline for approximately half the observations of NGC\,7412, increasing the noise and reducing our sensitivity to \mbox{H\,{\sc i}} absorption. 

In addition to the above observations, deeper follow-up observations were obtained for NGC\,7424. 
The primary aim of these observations was to resolve out the remaining \mbox{H\,{\sc i}} emission detected along this sightline and determine whether there was any \mbox{H\,{\sc i}} absorption present in the data. 
The follow-up observations were conducted over two periods of 11.5 and 10 hours on 2013 June 09 and 2013 June 11, respectively. 
These observations were conducted in the 6C array using the CABB 64M-32k configuration.
A summary of the ATCA observations is given in Table \ref{table:atca_obs}.

\begin{table}
\centering
\caption{Summary of the ATCA observations.}
\label{table:atca_obs}
\begin{threeparttable}
\begin{tabular}{@{} lllrl @{}} 
\hline
Target galaxy & Obs. Date & Array\tnote{$*$} & Int. time\\
 & & & (h) & \\
\hline
ESO\,150-G\,005 & 2011 Oct 27 & 750C & 8.10  \\
ESO\,345-G\,046 & 2011 Oct 28 & 750C & 8.36 \\
ESO\,402-G\,025 & 2012 Jul 08 & 750A & 8.17 \\
IC\,1954 & 2011 Nov 06 & 750C & 8.15 \\
NGC\,7412 & 2011 Oct 29 & 750C & 7.99 \\
NGC\,7424 & 2011 Oct 26 & 750C & 9.83 \\
Follow-up & 2013 Jun 09,11 & 6C & 14.02 \\
\hline
\end{tabular}
\begin{tablenotes}
\footnotesize{
\item[$*$] {Range of baseline lengths:}
\item[] {750A: 77-3750 m; 750C: 46-5020 m; 6C: 153-6000 m}
}
\end{tablenotes}
\end{threeparttable}
\end{table}

\subsection{Data reduction}
\label{data:data_reduction}

All data reduction was done using the {\sc miriad} package \citep{1995ASPC...77..433S}, implemented with a purpose-built automated data reduction pipeline, and following standard reduction procedures. 
For the follow-up observations of NGC\,7424, the data from the two epochs were calibrated separately and then combined to produce a single data-cube with a spectral resolution of 10 km s$^{-1}$.

We have produced 1.4 GHz continuum images and \mbox{H\,{\sc i}} data cubes for each of the targets in our sample. 
Spectra were produced towards each of the background continuum sources in our sample. 
In the case that a source was resolved into multiple components (see Figure \ref{figure:continuum_and_spectra}) a separate spectrum was produced for each component. 
All data-cubes have been corrected for primary beam effects.

We have produced three sets of data-cubes and continuum images, at different resolutions, which allow us to study both the \mbox{H\,{\sc i}} emission and absorption in the data. 
The three sets of images and cubes were produced using (i) uniform weighting, (ii) natural weighting \emph{including} the baselines to antenna 6, and (iii) natural weighting \emph{excluding} the baselines to antenna 6. 
Throughout this work the three versions are referred to as the high, medium, and low resolution cubes and images, respectively.

The three sets of cubes span the extremes of the resolution that can be achieved with the 750 m array (a few arcsec to about 1 arcmin). 
The lower resolution cubes are most sensitive to the diffuse \mbox{H\,{\sc i}} emission, allowing us to map the galaxies in \mbox{H\,{\sc i}}. 
\mbox{H\,{\sc i}} moment maps were produced from the low resolution data-cubes, to enable us to study the gas distribution and kinematics in the target galaxies.
At high resolution, all (or most) of the \mbox{H\,{\sc i}} emission is expected to have resolved out, allowing us to search for \mbox{H\,{\sc i}} absorption. 
The spectra produced from these cubes may be referred to as the `low/medium/high' resolution spectra, for short, which should be understood to refer to the spatial resolution of the cube (as opposed to the spectral resolution, which is always the same).

\section{Results I: H\,{\sevensize\bf I} distribution in the target galaxies}
\label{results_part1}

\subsection{H\,{\sevensize\bf I} emission maps}
\label{results_part1:hi_maps}

From our observations we have produced the first resolved \mbox{H\,{\sc i}} maps of the target galaxies. 
These maps (presented in Appendix \ref{appendix:moment_maps}) allow us to study the extent, distribution, and kinematics of the neutral gas. 
We find fairly symmetric, regular rotating disks with \mbox{H\,{\sc i}} masses of between $\sim$0.6 and 7 $\times$ 10$^{9}$ solar masses. 
Two galaxies, ESO\,402-G\,025 and NGC\,7412, show faint \mbox{H\,{\sc i}} extensions.

The \mbox{H\,{\sc i}} extension of ESO\,402-G\,025 is coincident with a known dwarf galaxy (GALEX source J212236.29-363701.8). 
The neighbouring galaxy, ESO\,402-G\,026, shows a strange `pinched' morphology and may have a slight extension towards ESO\,402-G\,025. 
Given the proximity of the galaxies, it seems likely that such features are the result of some past or on-going interaction.

The \mbox{H\,{\sc i}} masses obtained from our ATCA data are given in Table \ref{table:HI_masses} alongside the values from the HIPASS BGC \citep{2004AJ....128...16K}. 
As expected, our values are generally very close, but slightly lower than those obtained by HIPASS since an interferometer filters out the most diffuse \mbox{H\,{\sc i}} emission.

\begin{figure*}
\includegraphics[width=0.30\linewidth]{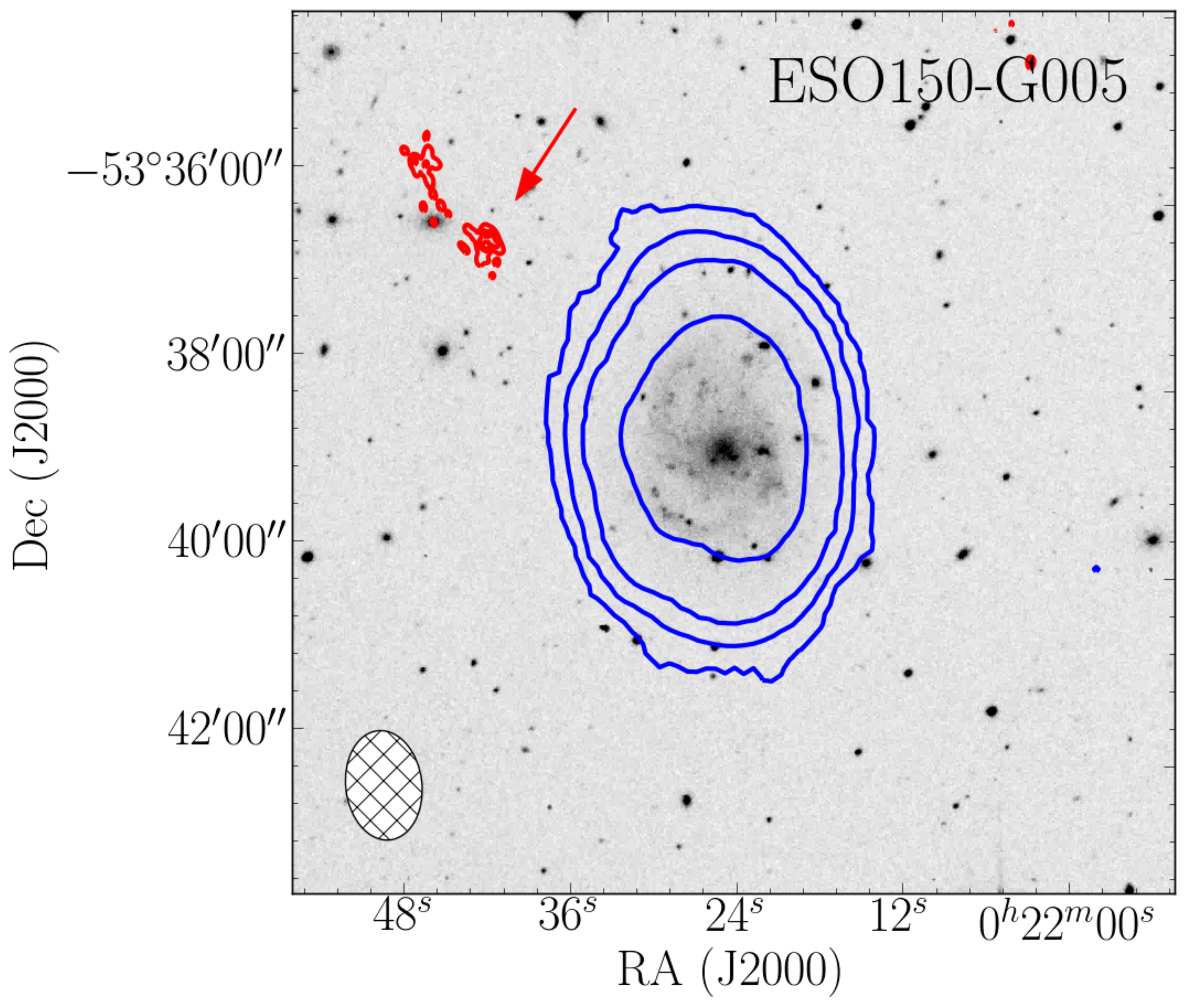}
\includegraphics[width=0.30\linewidth]{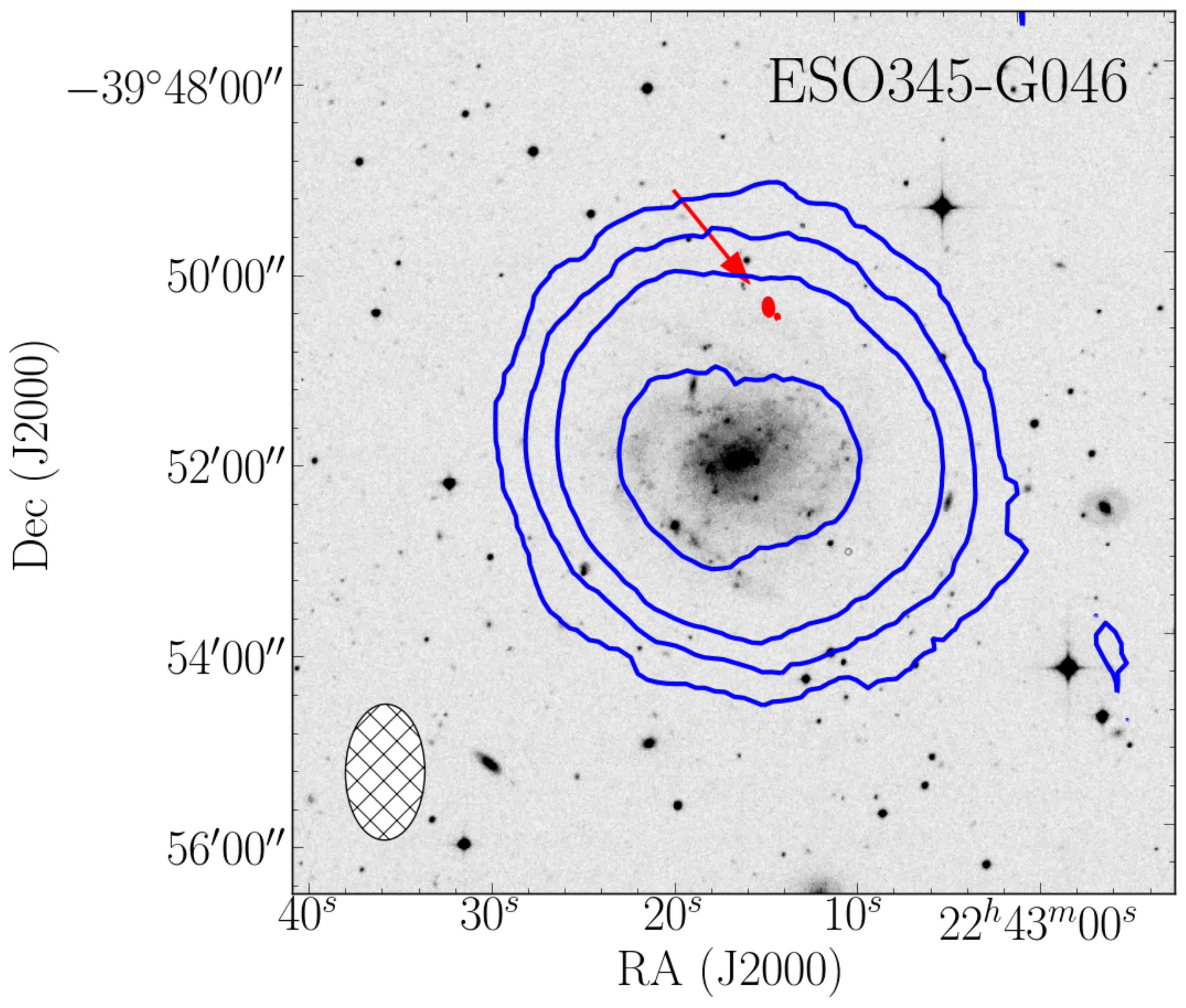} 
\includegraphics[width=0.30\linewidth]{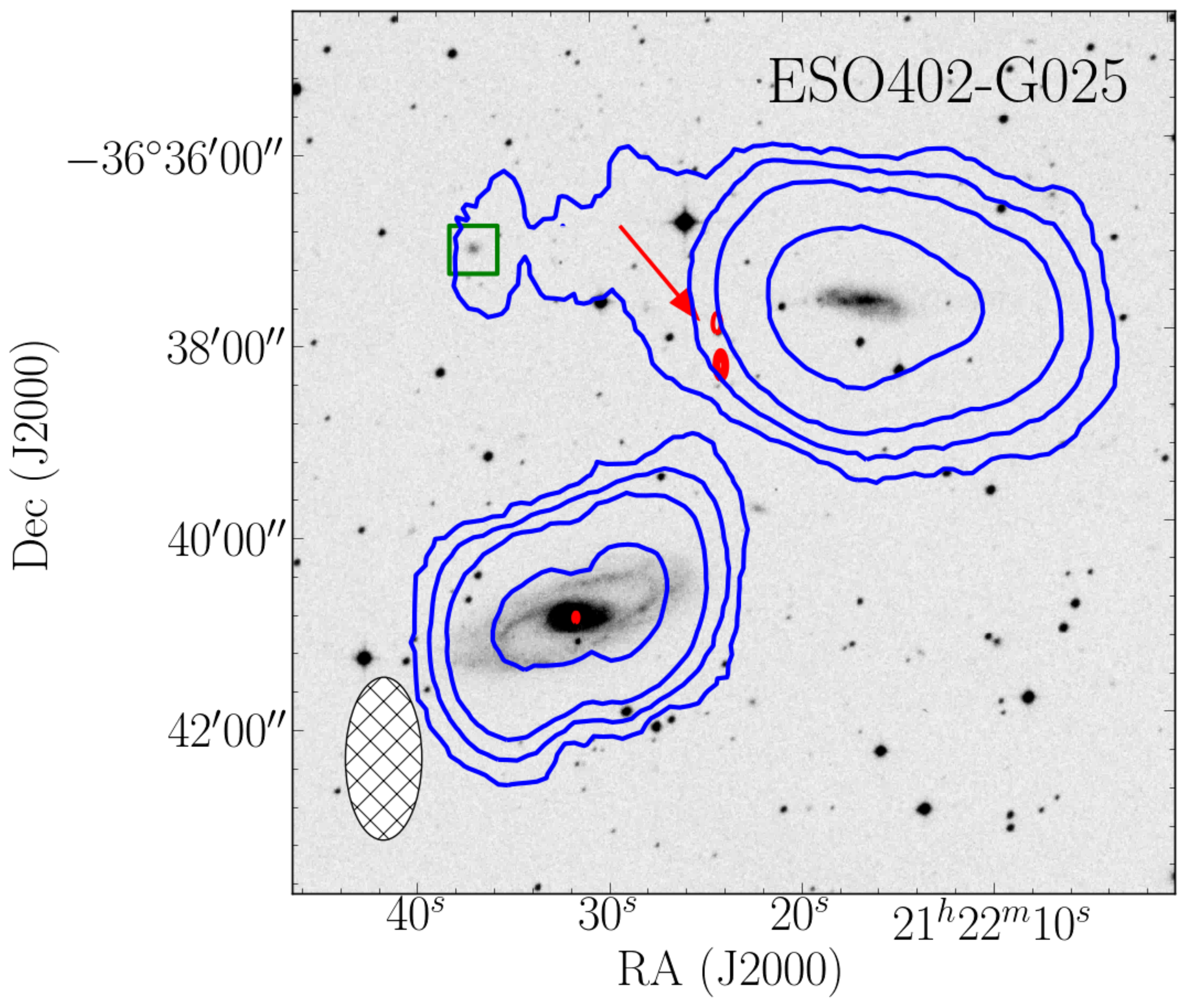}
\includegraphics[width=0.30\linewidth]{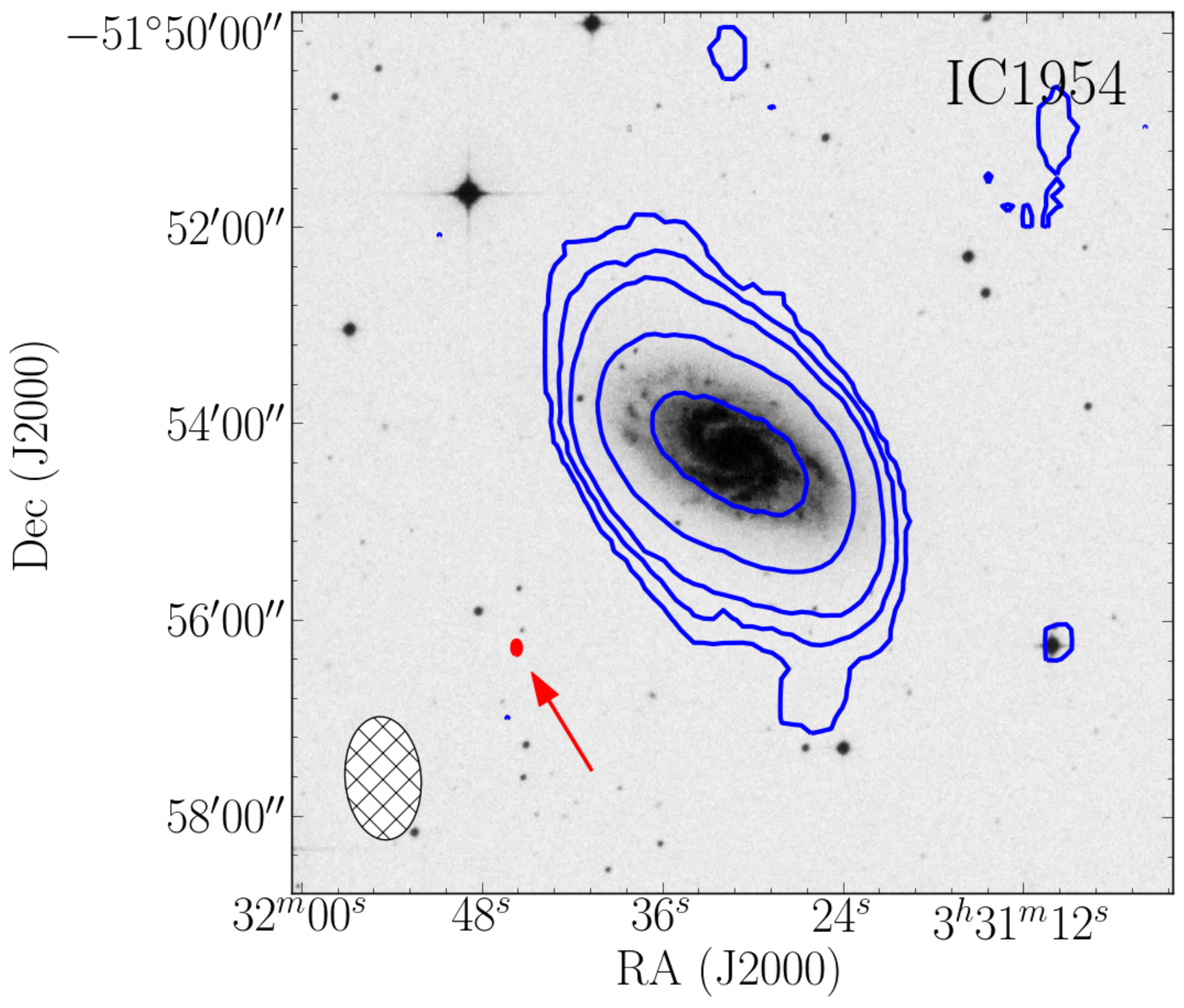}
\includegraphics[width=0.30\linewidth]{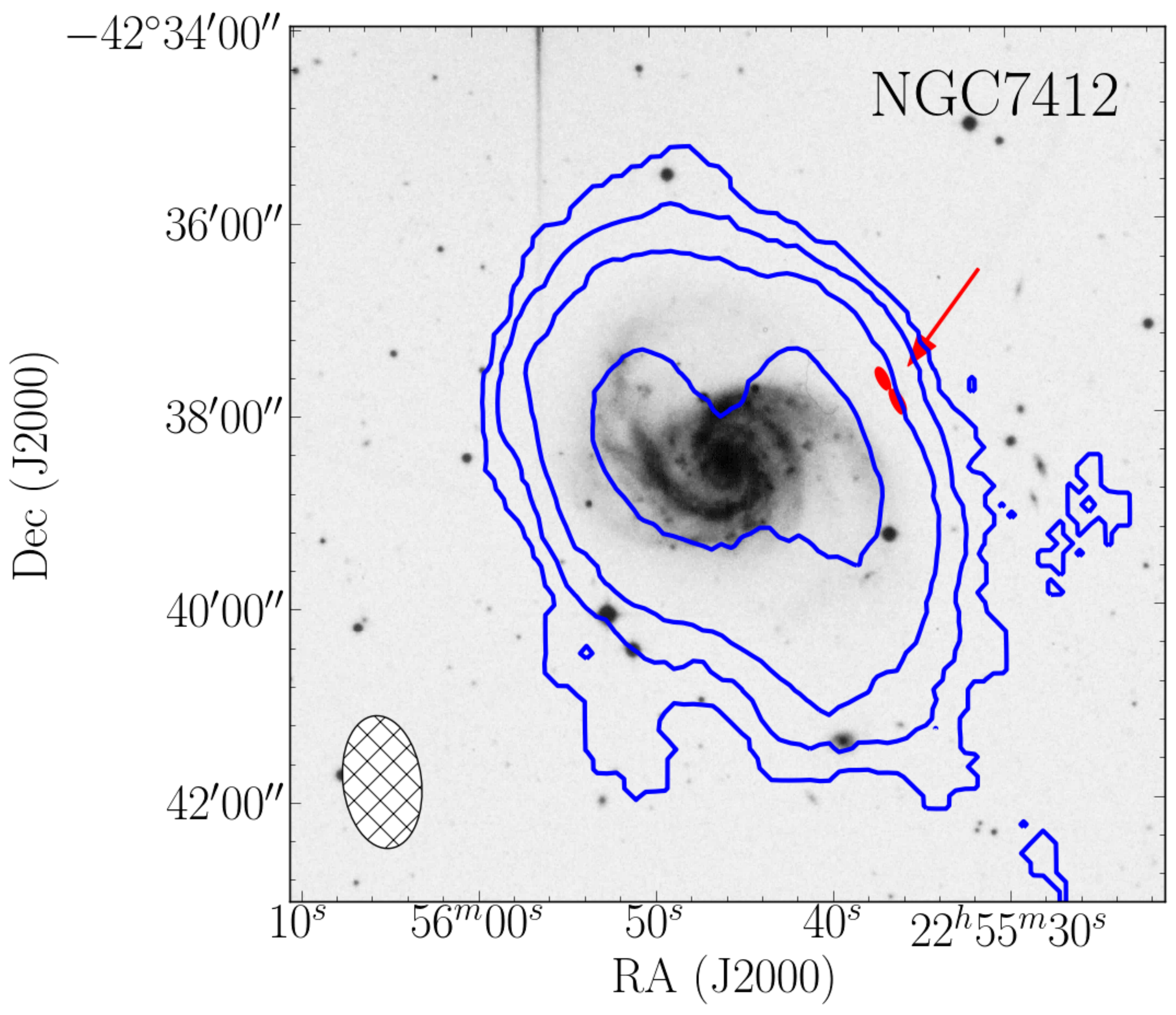}
\includegraphics[width=0.30\linewidth]{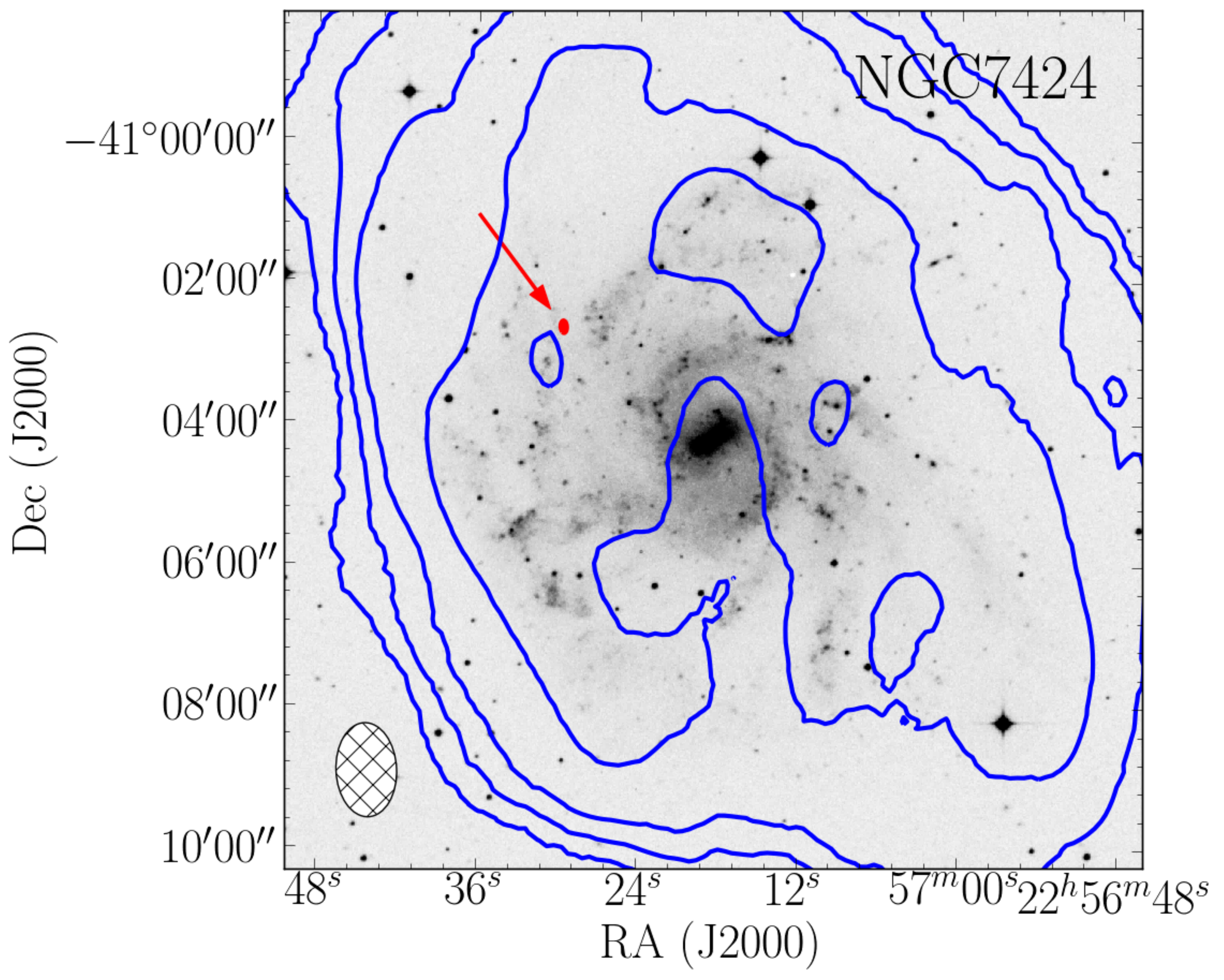}
\caption[]{SuperCOSMOS B-band images \citep{2001MNRAS.326.1279H} overlaid with \mbox{H\,{\sc i}} contours (blue) overlaid. 
Contour levels are $N_{\mathrm{HI}}$ = 3 $\times$ 10$^{19}$, (1, 2, 5) $\times$ 10$^{20}$, and 1 $\times$ 10$^{21}$ cm$^{-2}$. 
The high resolution 1.4 GHz radio continuum emission is also overlaid (red contours), with an arrow indicating the location of the background continuum source(s). 
The location of the dwarf galaxy near ESO\,402-G\,025 is marked by the green box. 
The synthesised beam for the \mbox{H\,{\sc i}} maps is shown in the bottom left corner.}
\label{figure:overlay_maps}
\end{figure*}

In Figure \ref{figure:overlay_maps} we show the \mbox{H\,{\sc i}} emission overlaid on the optical images. 
The high resolution 1.4 GHz radio continuum emission is also shown, indicating the position of the background source(s) and we find that four of the sightlines intersect the \mbox{H\,{\sc i}} disks at column densities of above $\sim$1-2 $\times$ 10$^{20}$ cm$^{-2}$.

\begin{table}
\centering
\caption{Comparison of the HIPASS and ATCA \mbox{H\,{\sc i}} masses of the target galaxies, assuming $D = v_{\mathrm{LG}}/H_{0}$. 
The HIPASS masses have been re-scaled using a Hubble constant of H$_{0}$ = 71 km s$^{-1}$ Mpc$^{-1}$, as elsewhere in this work. 
For the galaxy pair ESO\,402-G\,025/6 (HIPASS J2122-36) we give the combined mass of the two galaxies.}
\label{table:HI_masses}
\begin{tabular}{@{} lrrrr @{}} 
\hline
& \multicolumn{2}{c}{logM$_{\mathrm{HI}}$ (M$\odot$)} \\
\hline
Galaxy & This work & HIPASS \\
\hline
ESO\,150-G\,005 & 9.03 & 9.05 \\
ESO\,345-G\,046 & 9.58 & 9.73 \\
ESO\,402-G\,025/6 & 9.77 & 9.71 \\
IC\,1954 & 8.75 & 8.85 \\
NGC\,7412 & 9.46 & 9.58 \\
NGC\,7424 & 9.81 & 9.96 \\
\hline
\end{tabular}
\end{table}

\subsection{Radial H\,{\sevensize\bf I} profiles}
\label{results_part1:radial_profiles}

\begin{figure*}
\includegraphics[width=\textwidth]{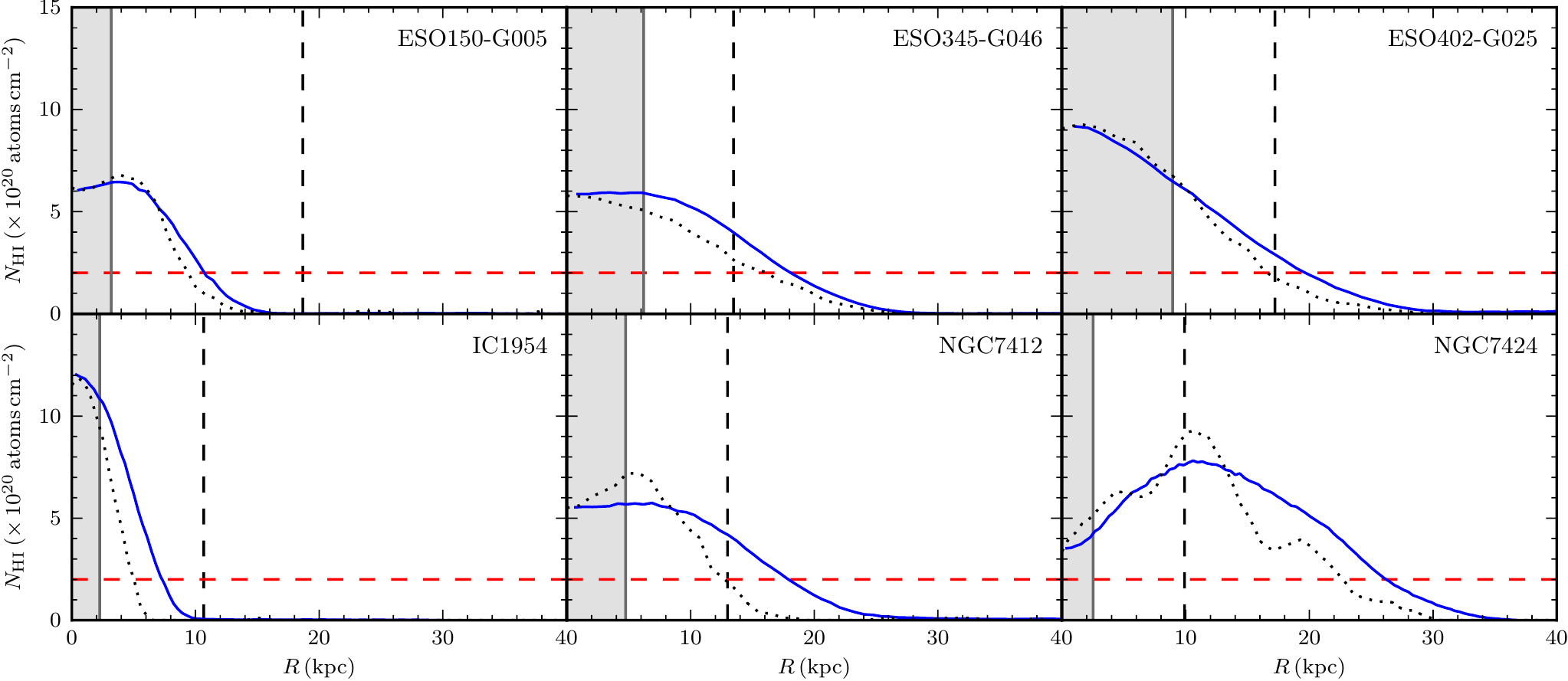}
\caption{Radial \mbox{H\,{\sc i}} profiles of the galaxies in our sample, derived from the ATCA \mbox{H\,{\sc i}} total intensity maps. 
The blue (solid) curves show the azimuthally averaged profile, and the black (dotted) curves the profile along the axis towards the continuum source. 
The impact parameter of the continuum source is shown by the dashed vertical line (for ESO\,150-G\,005 we use component 1, the nearer of the two components). 
The grey shaded region indicates semi-major axis of the synthesised beam and dashed horizontal line indicates the DLA limit ($N_{\mathrm{HI}}$ = 2 $\times$ 10$^{20}$ cm$^{-2}$).}
\label{figure:radial_profiles}
\end{figure*}

In Figure \ref{figure:radial_profiles} we show radial \mbox{H\,{\sc i}} profiles produced from the \mbox{H\,{\sc i}} maps of the target galaxies. 
For each galaxy we show two profiles -- the azimuthally averaged profile, and the profile along the axis towards the continuum source.
The ellipse parameters used in producing the azimuthally averaged profiles are given in Table \ref{table:ellint_parameters}.

These profiles provide us with additional information about the extent of the \mbox{H\,{\sc i}} disks, and provides a complement to the absorption-line data, which can help in estimating the likely detection rate as a function of impact parameter (see Section \ref{discussion:detection_rate_impact_parameter}).
At $N_{\mathrm{HI}}$ = 2 $\times$ 10$^{20}$ cm$^{-2}$ the observed \mbox{H\,{\sc i}} disks are about 50 per cent larger than the stellar disks at a B-band surface brightness of 25 mag arcsec$^{-2}$ (see Table \ref{table:HI_opt_disk_sizes}), except for ESO\,402-G\,025, which has an \mbox{H\,{\sc i}} to stellar disk ratio of 3.4. 
This is likely due to the eastern extension and companion galaxy as discussed in Section \ref{results_part1:hi_maps}.

Comparison of the azimuthal and continuum profiles shows that, in general there is not a dramatic difference between the two, with the biggest differences seen in the galaxies that are either highly inclined or highly asymmetric. 
This means that for blind absorption-line surveys, even if the continuum source is not located along the major axis, the likelihood of detecting intervening absorption at a certain impact parameter is probably not much lower.

\begin{table}
\centering
\caption{Ellipse parameters, derived from the \mbox{H\,{\sc i}} maps, used to produce the azimuthally averaged radial profiles shown in Figure \ref{figure:radial_profiles}.}
\label{table:ellint_parameters}
\begin{tabular}{@{} lrr @{}} 
\hline
Galaxy & Ellipticity & Position Angle \\
&  & (deg) \\
\hline
ESO\,150-G\,005 & 0.30 & 5 \\
ESO\,345-G\,046 & 0.08 & 51 \\
ESO\,402-G\,025 & 0.31 & 79 \\
IC\,1954 & 0.38 & 41 \\
NGC\,7412 & 0.30 & 32 \\
NGC\,7424 & 0.28 & 32 \\
\hline
\end{tabular}
\end{table}

\begin{table}
\centering
\caption{Comparison of \mbox{H\,{\sc i}} and optical disk sizes for the galaxies in our sample.}
\label{table:HI_opt_disk_sizes}
\begin{threeparttable}
\begin{tabular}{@{} lrrr @{}} 
\hline
Galaxy & R$_{\mathrm{opt}}$ & R$_{\mathrm{HI}}$ & R$_{\mathrm{HI}}$/R$_{\mathrm{opt}}$ \\
 & (kpc) & (kpc) & \\
\hline
ESO\,150-G\,005 & 7.3\tnote{$1$} & 11.7 & 1.6 \\
ESO\,345-G\,046 & 11.2\tnote{$1$} & 18.3 & 1.6 \\
ESO\,402-G\,025 & 5.8\tnote{$2*$} & 19.8 & 3.4 \\
IC\,1954 & 5.7\tnote{$1$} & 9.1 & 1.6 \\
NGC\,7412 & 13.4\tnote{$1$} & 18.8 & 1.4 \\
NGC\,7424 & 18.1\tnote{$1$} & 26.8 & 1.5 \\
\hline
\end{tabular}
\begin{tablenotes}
\footnotesize{
\item[] {Optical diameter references:}
\item[$1$] RC3 {\citep{1991rc3..book.....D}}
\item [$2$] ESO {\citep{1982euse.book.....L}}
\item [$*$] For ESO\,402-G\,025 no $D_{25}$ measurement was available, so we have instead used the closest available measurement, which is the diameter at a B-band surface brightness of about 25.5 mag arcsec$^{-2}$}
\end{tablenotes}
\end{threeparttable}
\end{table}

\section{Results II: Spectra towards the background radio sources}
\label{results_part2}

\subsection{Spectra and continuum images}
\label{results_part2:continuum_images_and_spectra}

Figure \ref{figure:continuum_and_spectra} shows the 1.4 GHz ATCA continuum images of the background sources used to search for absorption, and below each the spectrum towards that source. 
For double sources the individual spectra are shown one above the other, brightest component first.

We applied a Bayesian analysis to all spectra, using the {\sc multi-nest} algorithm \citep{2008MNRAS.384..449F,2009MNRAS.398.1601F}, to determine whether a spectral-line was present in the data and, if so, the best-fitting parameters for the line. 
For a full description of the development, testing and implementation of this algorithm for spectral-line data we refer the reader to \citet{2012PASA...29..221A,2012MNRAS.423.2601A,2013MNRAS.430..157A}.

We define the quantity 
\begin{equation}
R = \text{ln}\left( \frac{E_{\mathrm{line}}}{E_{\mathrm{noise}}} \right),
\label{equation:R_parameter}
\end{equation}
which is the ratio of the `Bayesian Evidence' for a noise-only model ($E_{\mathrm{noise}}$) and a noise plus spectral-line model ($E_{\mathrm{line}}$). 
Values of $R$ less than zero indicate that the data 
do not warrant the inclusion of a spectral line component in the model, while values of $R$ greater than zero indicate that the inclusion of a spectral-line is warranted (the larger the value of $R$ the stronger the evidence for the detection).

\begin{figure*}
\includegraphics[width=\linewidth]{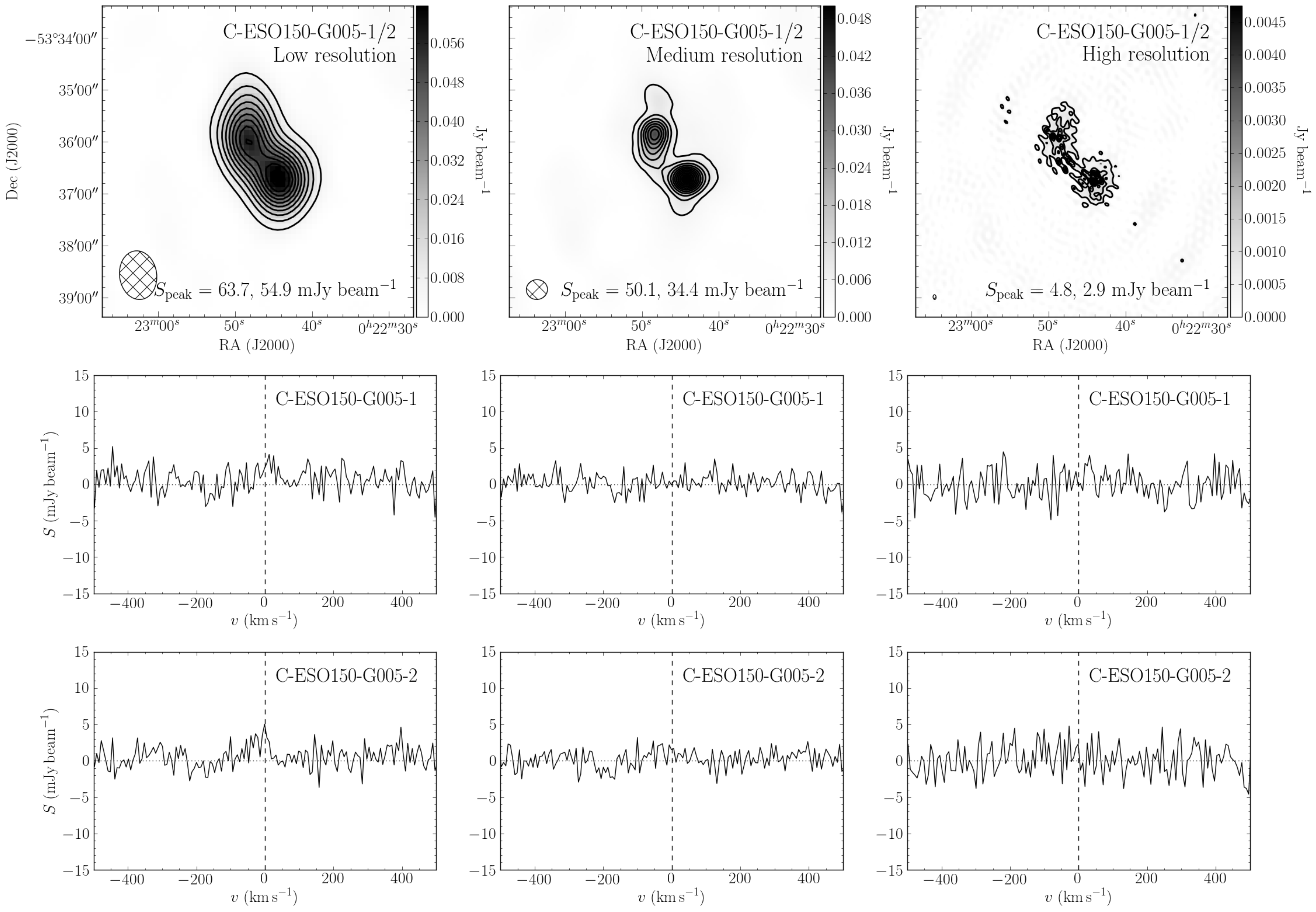}
\includegraphics[width=\linewidth]{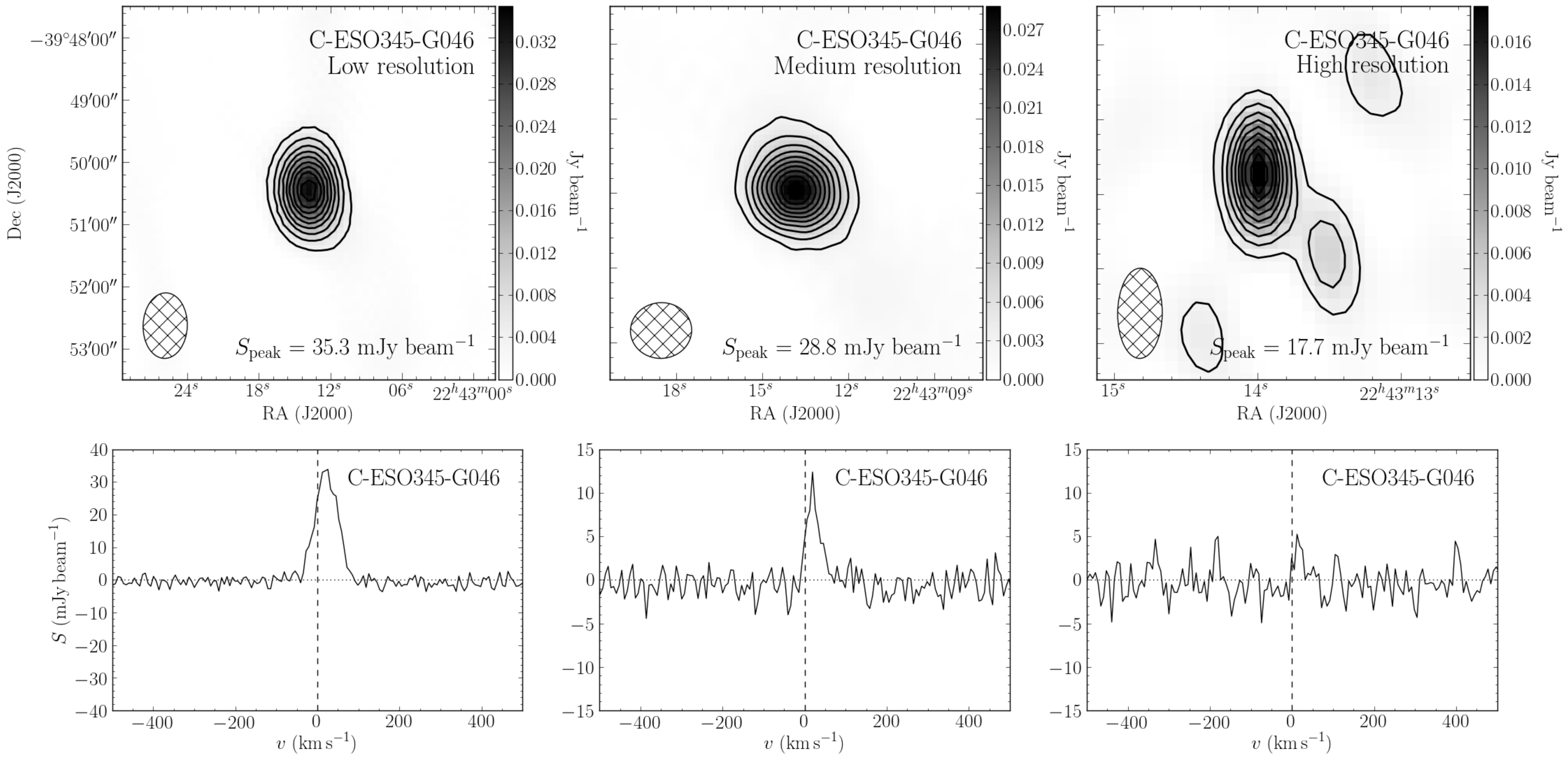}
\caption[]{ATCA 1.4 GHz continuum images of the background radio sources, with the spectrum along that sightline below it (left to right: low, medium, and high resolution). 
All spectra have been shifted to the rest-frame of the galaxy, with the velocity axis expressed with respect to the galaxy systemic velocity. 
The dotted line represents the approximate expected 21 cm line position. 
We note that, apart from C-ESO\,150-G\,005-1/2, the continuum images at higher resolution show a smaller region of sky in order to see the source structure. 
The radio contours start at 90 per cent of the peak flux and decrease in 10 per cent increments, and the peak flux is given in the bottom right corner. 
The synthesised beam is shown in the bottom left corner of each continuum image.}
\label{figure:continuum_and_spectra}
\end{figure*}

\begin{figure*}
\includegraphics[width=\linewidth]{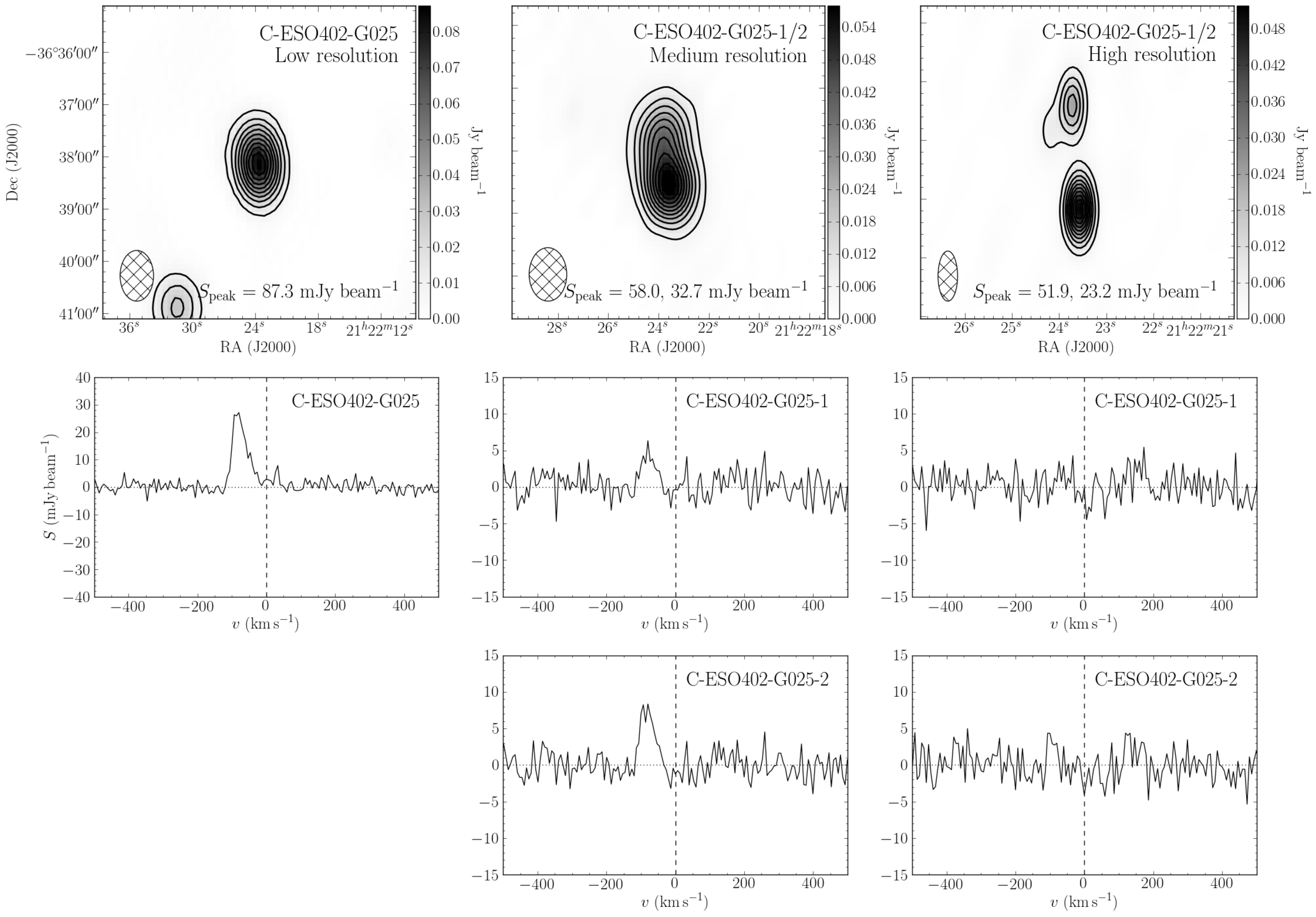}
\includegraphics[width=\linewidth]{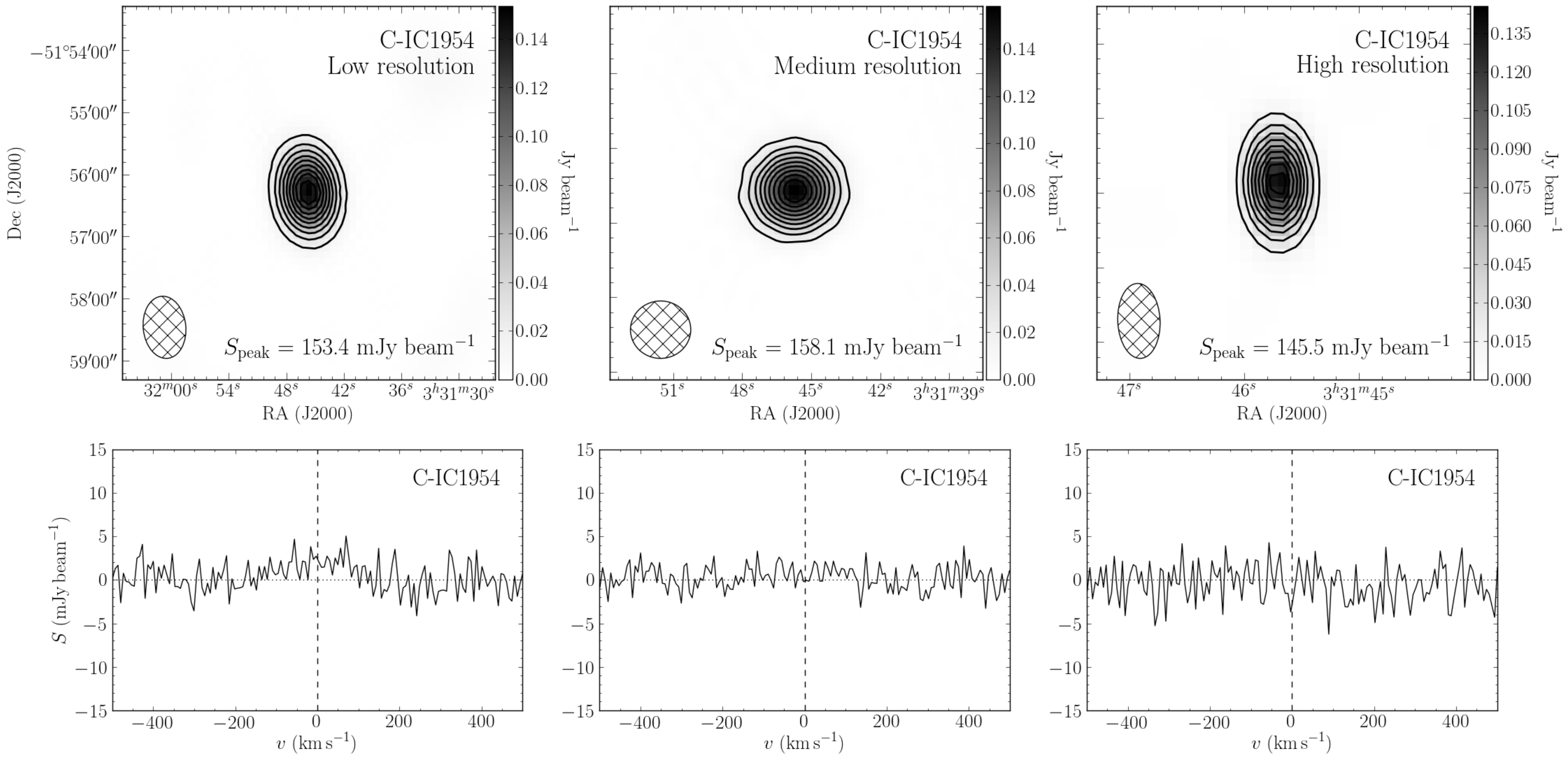}
\contcaption{}
\end{figure*}

\begin{figure*}
\includegraphics[width=\linewidth]{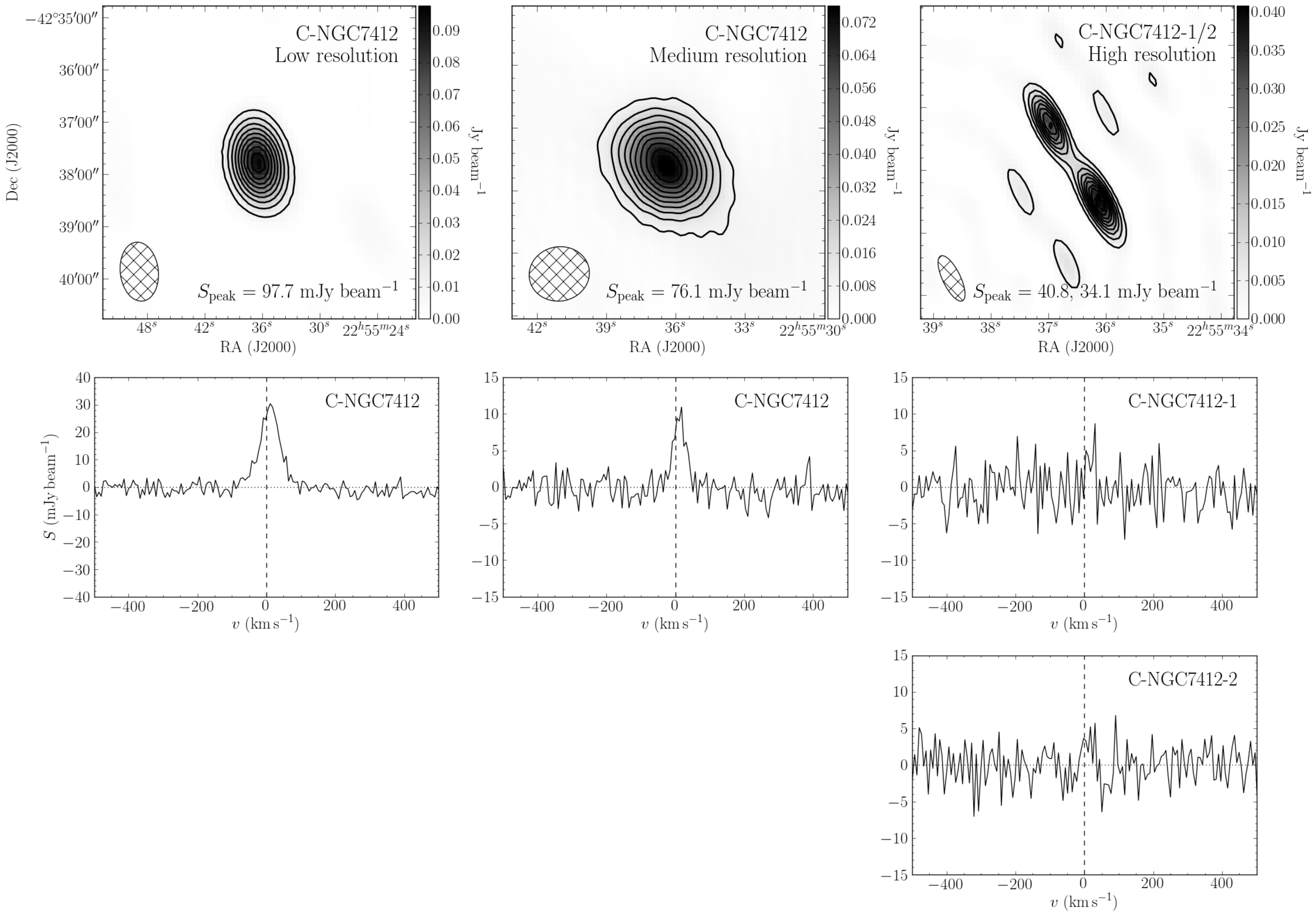}
\includegraphics[width=\linewidth]{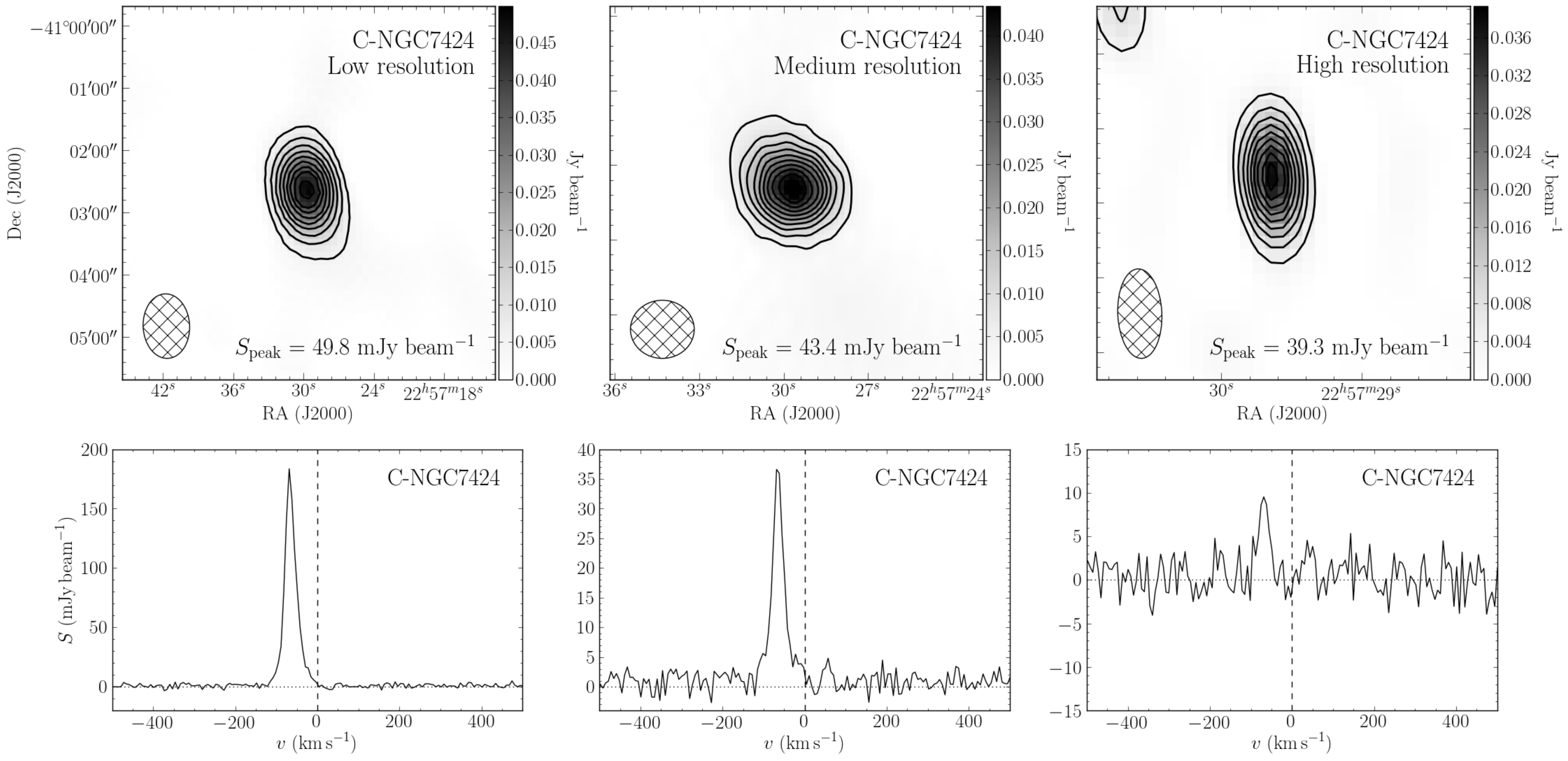}
\contcaption{}
\end{figure*}

\subsection{Sightlines with H\,{\sevensize\bf I} emission}
\label{results_part2:hi_emission}

Emission-lines were detected along sightlines in four of the six target galaxies -- ESO\,345-G\,046, ESO\,402-G\,025, NGC\,7412, and NGC\,7424 (all of which clearly intersect the \mbox{H\,{\sc i}} disk in Figure \ref{figure:overlay_maps}). 
By far the strongest detection is along the sightline towards C-NGC\,7424, which has a peak line flux of almost 200 mJy beam$^{-1}$ in the low resolution spectrum.

As expected, the detected emission-lines are strongest in the low resolution spectra and decrease in strength with increasing resolution.
Detections were made in the low, medium, and high resolution spectra towards C-ESO\,345-G\,046 and C-NGC\,7424, and in the low and medium resolution spectra towards C-ESO\,402-G\,025 and C-NGC\,7412. 
Visually, there also appears to be weak emission in the high resolution spectra towards C-NGC\,7412-1/2, at a velocity consistent with the lower resolution detections, but the noise in the spectra is too high to be able to confirm this at present.

We note that the small peaks seen in the low resolution spectra towards C-ESO\,150-G\,005-1/2 and C-IC\,1954, close to the systemic velocity, are actually artefacts resulting from emission closer to the galaxy which is convolved with the side-lobes of the synthesised beam. 
We also searched the spectrum towards C-ESO\,402-G\,025 for emission from the companion galaxy ESO\,402-G\,026 but, as can be seen in Figure \ref{figure:overlay_maps}, the companion is too distant, and we did not detect emission in the spectra at any resolution. 

We have calculated the \mbox{H\,{\sc i}} column density along each of the sightlines where emission was detected, and find it to be typically a few times 10$^{20}$ cm$^{-2}$, although significantly higher in the case of NGC\,7424. 
The parameters of each of the detected lines, and the calculated column densities, are given in Table \ref{table:emission_line_results}.

\begin{table*}
\begin{minipage}{\linewidth}
\centering
\caption{Best-fitting parameters estimated for the detected \mbox{H\,{\sc i}} emission-lines. 
Columns (1) and (2) are the sightline and cube resolution at which the line was detected. 
Columns (3), (4), and (5) are the radial velocity, width, and peak flux of the line. 
The line width is defined as the Full Width at Half Maximum (FWHM). 
Column (6) is the integrated \mbox{H\,{\sc i}} line flux. 
Column (7) is the derived column density. 
Column (8) is the $R$-value of the detected line (defined in Equation \ref{equation:R_parameter}).}
\label{table:emission_line_results}
\begin{tabular}{@{} llrrrrrr @{}} 
\hline
Sightline & Cube & Velocity ($cz$) & S$_{\mathrm{peak}}$ & Width ($\Delta v$) & $\int S\,dv$ & N$_{\mathrm{HI}}$ & R-value \\
 & resolution & (km s$^{-1}$) & (mJy bm$^{-1}$) & (km s$^{-1}$) & (mJy bm$^{-1}$ km s$^{-1}$) & ($\times$10$^{20}$ cm$^{-2}$) &  \\
\hline
C-ESO\,345-G\,046 & Low & 2171.1$^{+0.6}_{-0.6}$ & 34.5$^{+0.7}_{-0.7}$ & 65.8$^{+1.6}_{-1.5}$ & 2272.7$^{+45.8}_{-45.0}$ & 6.0$^{+0.1}_{-0.1}$ & 1757.89$ \pm $0.07 \\
C-ESO\,345-G\,046 & Medium & 2169.2$^{+1.6}_{-1.6}$ & 10.6$^{+1.0}_{-1.0}$ & 36.6$^{+4.1}_{-3.8}$ & 387.8$^{+34.3}_{-33.3}$ & 7.8$^{+0.7}_{-0.7}$ & 89.21$ \pm $0.06 \\
\vspace{+1mm}
C-ESO\,345-G\,046 & High & 2165.0$^{+4.0}_{-2.8}$ & 5.1$^{+4.7}_{-2.1}$ & 17.8$^{+11.1}_{-14.1}$ & 86.1$^{+35.2}_{-36.8}$ & 28.4$^{+11.6}_{-12.2}$ & 1.86$ \pm $0.04 \\
C-ESO\,402-G\,025 & Low & 2494.2$^{+1.0}_{-0.9}$ & 27.6$^{+1.1}_{-1.1}$ & 49.2$^{+2.4}_{-2.3}$ & 1358.9$^{+51.3}_{-50.9}$ & 3.1$^{+0.1}_{-0.1}$ & 527.18$ \pm $0.06 \\
C-ESO\,402-G\,025-1 & Medium & 2493.5$^{+4.5}_{-4.6}$ & 4.8$^{+1.1}_{-1.0}$ & 44.4$^{+10.5}_{-9.3}$ & 212.5$^{+46.1}_{-43.1}$ & 5.2$^{+1.1}_{-1.1}$ & 10.28$ \pm $0.05 \\
\vspace{+1mm}
C-ESO\,402-G\,025-2 & Medium & 2490.3$^{+2.6}_{-2.6}$ & 8.3$^{+1.0}_{-0.9}$ & 45.7$^{+5.8}_{-5.2}$ & 379.6$^{+42.0}_{-41.6}$ & 9.3$^{+1.0}_{-1.0}$ & 46.95$ \pm $0.06 \\
C-NGC\,7412 & Low & 1720.7$^{+1.1}_{-1.1}$ & 29.5$^{+1.0}_{-1.0}$ & 70.9$^{+3.0}_{-2.9}$ & 2096.1$^{+70.0}_{-70.4}$ & 5.8$^{+0.2}_{-0.2}$ & 705.49$ \pm $0.07 \\
\vspace{+1mm}
C-NGC\,7412 & Medium & 1721.9$^{+2.1}_{-2.2}$ & 10.1$^{+1.1}_{-1.0}$ & 44.1$^{+5.6}_{-5.0}$ & 443.5$^{+44.2}_{-43.0}$ & 6.6$^{+0.7}_{-0.6}$ & 65.59$ \pm $0.06 \\
C-NGC\,7424 & Low & 872.9$^{+0.1}_{-0.1}$ & 181.3$^{+1.4}_{-1.3}$ & 33.4$^{+0.3}_{-0.3}$ & 6058.9$^{+42.6}_{-42.9}$ & 16.5$^{+0.1}_{-0.1}$ & 15620.70$ \pm $0.08 \\
C-NGC\,7424 & Medium & 873.8$^{+0.5}_{-0.5}$ & 37.0$^{+1.3}_{-1.3}$ & 35.3$^{+1.6}_{-1.5}$ & 1303.9$^{+42.2}_{-41.8}$ & 24.0$^{+0.8}_{-0.8}$ & 807.05$ \pm $0.07 \\
C-NGC\,7424 & High & 869.8$^{+2.1}_{-2.2}$ & 9.8$^{+1.5}_{-1.4}$ & 32.0$^{+6.1}_{-5.2}$ & 314.5$^{+44.0}_{-42.4}$ & 102.6$^{+14.3}_{-13.8}$ & 31.77$ \pm $0.05 \\
\hline
\end{tabular}
\end{minipage}
\end{table*}

\subsection{Search for H\,{\sevensize\bf I} absorption}
\label{results_part2:hi_absorption}

We have searched for \mbox{H\,{\sc i}} absorption in all of the high resolution spectra but have not detected any absorption-lines (with the exception of Galactic absorption along the sightlines to C-ESO\,150-G\,005-1/2). 
The follow-up observations of NGC\,7424 confirm that there is no intervening absorption seen in this galaxy. 
We also searched for associated absorption against the radio source located in ESO\,402-G\,026, but found no evidence of absorption.

In Table \ref{table:absorption_line_results} we give the 3-$\sigma$ upper limits on the optical depth and \mbox{H\,{\sc i}} column density. 
All limits are calculated for a spin temperature of $T_{\mathrm{S}}$ = 100 K, and a covering factor of $f$ = 1. 
We assume a gaussian line-profile with a FWHM of 10 km s$^{-1}$. 
These assumptions are the same as used by \citet{2010MNRAS.408..849G}, 
allowing us to directly compare our results (see Section \ref{figure:gupta_comparison}).

\subsection{Limits on the spin temperature}
\label{results_part2:spin_temp}

For the sightlines where both \mbox{H\,{\sc i}} emission and \mbox{H\,{\sc i}} absorption are detected it is possible to obtain a good estimate for the ratio of the spin temperature and the covering factor ($T_{\mathrm{S}}/f$), which is not possible with absorption-line data alone. 
Although we haven't detected any absorption-lines in our sample, we can still place lower limits on the value of $T_{\mathrm{S}}/f$. 
The ratio $T_{\mathrm{S}}/f$ is given by:
\begin{equation}
\frac{T_{\mathrm{S}}}{f} \approx \frac{N_{\mathrm{HI,em}}}{\displaystyle 1.823 \times 10^{18} \int \tau_{\mathrm{obs}} \ dv}, 
\label{equation:spin_temp}
\end{equation}
where $N_{\mathrm{HI,em}}$ is the \mbox{H\,{\sc i}} column density along that sightline, calculated from the emission-line data, $\tau_{\mathrm{obs}}$ is the observed optical depth. 

We have calculated lower limits for $T_{\mathrm{S}}/f$ along each of the sightlines where an \mbox{H\,{\sc i}} emission-line was detected, and these are presented in the final column of Table \ref{table:absorption_line_results}. 
In most cases, the limits we obtain are around 100-200 K, and for galaxies with multiple sightlines the individual sightlines yield similar limits.
These values are in good agreement with \citet{1992ApJ...399..373C} who found lower limits of a few hundred K from similar emission- and absorption-line observations.
\citet{2010ApJ...713..131B,2014ApJ...795...98B} and \citet{2013MNRAS.428.2198S} also estimate spin temperatures of a few tens to a few hundred K, based on line-widths and column density measurements from Lyman-$\alpha$ absorption.

However, for NGC\,7424, we obtain a much higher limit for $T_{\mathrm{S}}/f$. 
The original observations gave a limit of $T_{\mathrm{S}}/f$ $\gtrsim$ 3600 K, but with the lower noise in the follow-up observations, we calculate a revised limit of $T_{\mathrm{S}}/f$ $\gtrsim$ 5700 K.
We suggest that the reason for this high value is a clumpy \mbox{H\,{\sc i}} medium, in which small, dense clouds are embedded in a more diffuse medium. 
This would result in a high averaged column density in the larger \mbox{H\,{\sc i}} emission beam, as observed, but could also explain the absorption-line non-detection if the narrow quasar sightline misses these dense clumps. 
In this scenario, although $T_{\mathrm{S}}/f$ is very high, the actual spin temperature may not be unusually high if the covering factor is very low, due to the clumpiness of the gas.

Such structures have previously been suggested by \citet{2012ApJ...749...87B}. 
\citet{2013MNRAS.428.2198S,2010ApJ...713..131B,2011ApJ...727...52B,2014ApJ...795...98B} have all recently inferred the presence of dense, parsec-scale \mbox{H\,{\sc i}} clouds from \mbox{H\,{\sc i}} absorption-line observations. 
In addition, VLBI studies have shown that the detection rate of \mbox{H\,{\sc i}} is highest for background quasars with linear sizes less than 100 pc, suggesting that this also the typical correlation length of \mbox{H\,{\sc i}} clouds \citep{2012A&A...544A..21G}. 
With present observations we are unable to determine whether this is the case for NGC\,7424, as we are not able to resolve structures on scales smaller than about 300 pc, however future observations, for example using VLBI spectroscopy, would enable us to better understand the nature of the gas in this galaxy.

\begin{table*}
\begin{minipage}{\linewidth}
\centering
\caption{Absorption-line upper limits and limits on $T_{\mathrm{S}}/f$. 
Column (1) is the sightline searched. 
Columns (2)-(6) are the values derived from the absorption-line data (3-$\sigma$ upper limits for the non-detections). 
Column (2) is the rms-noise level. 
Column (3) is the peak 1.4 GHz continuum flux of the background source. 
Column (4) is the peak optical depth of the line.
Column (5) is the integrated optical depth, assuming a line-width of 10 km s$^{-1}$. 
Column (6) is the absorption-line \mbox{H\,{\sc i}} column density. 
Columns (7) and (8) are the relevant emission-line values used to estimate $T_{\mathrm{S}}/f$.
Column (7) is the emission-line column density. 
Column (8) is the weighting scheme from which that column density was derived. 
Column (9) is the limit for $T_{\mathrm{S}}/f$ calculated from the combined emission- and absorption-line data. 
If an emission-line was detected at multiple spatial resolutions, we have calculated a separate limit for $T_{\mathrm{S}}/f$ at each resolution. 
The June 2013 limit for NGC\,7424 uses the emission-line column density from the October 2011 observation, which is denoted by square brackets.}
\label{table:absorption_line_results}
\begin{tabular}{@{} lrrrrrrlr @{}} 
\hline
&  \multicolumn{5}{l}{Absorption-line (high resolution spectra)} & \multicolumn{2}{l}{Emission-line} & \multicolumn{1}{l}{Combined}\\
\hline
Sightline & $\sigma_{\mathrm{chan}}$ & S$_{\mathrm{peak,1.4}}$ & $\tau_{\mathrm{peak}}$ & $\int \tau\,dv$ & N$_{\mathrm{HI}}$(abs) & N$_{\mathrm{HI}}$(em) & Cube & T$_{\mathrm{S}}/f$ \\
 & (mJy bm$^{-1}$) & (mJy bm$^{-1}$) & (per cent) & (km s$^{-1}$) & ($\times$10$^{20}$ cm$^{-2}$) & ($\times$10$^{20}$ cm$^{-2}$) & resolution & (K) \\
\hline
C-ESO\,150-G\,005-1 & 2.11 & 4.8* & $<$1.33 & $<$14.08 & $<$25.7 & - & - & - \\
\vspace{+1.00mm}
C-ESO\,150-G\,005-2 & 2.13 & 2.9* & $<$2.20 & $<$23.32 & $<$42.5 & - & - & - \\
C-ESO\,345-G\,046 & 1.62 & 17.7$ \pm $1.7 & $<$0.27 & $<$2.90 & $<$5.3 & 6.0$^{+0.1}_{-0.1}$ & Low & $>$113 \\
C-ESO\,345-G\,046 & 1.62 & 17.7$ \pm $1.7 & $<$0.27 & $<$2.90 & $<$5.3 & 7.8$^{+0.7}_{-0.7}$ & Medium & $>$147 \\
\vspace{+1.00mm}
C-ESO\,345-G\,046 & 1.62 & 17.7$ \pm $1.7 & $<$0.27 & $<$2.90 & $<$5.3 & 28.4$^{+11.6}_{-12.2}$ & High & $>$537 \\
C-ESO\,402-G\,025-1 & 1.93 & 51.9$ \pm $0.8 & $<$0.11 & $<$1.18 & $<$2.2 & 3.1$^{+0.1}_{-0.1}$ & Low & $>$146 \\
C-ESO\,402-G\,025-1 & 1.93 & 51.9$ \pm $0.8 & $<$0.11 & $<$1.18 & $<$2.2 & 5.2$^{+1.1}_{-1.1}$ & Medium & $>$241 \\
C-ESO\,402-G\,025-2 & 2.11 & 23.2$ \pm $1.2 & $<$0.27 & $<$2.88 & $<$5.3 & 3.1$^{+0.1}_{-0.1}$ & Low & $>$60 \\
\vspace{+1.00mm}
C-ESO\,402-G\,025-2 & 2.11 & 23.2$ \pm $1.2 & $<$0.27 & $<$2.88 & $<$5.3 & 9.3$^{+1.0}_{-1.0}$ & Medium & $>$177 \\
\vspace{+1.00mm}
C-IC\,1954 & 2.19 & 145.5$ \pm $1.0 & $<$0.05 & $<$0.48 & $<$0.9 & - & - & - \\
C-NGC\,7412-1 & 2.67 & 40.8$ \pm $3.4 & $<$0.20 & $<$2.08 & $<$3.8 & 5.8$^{+0.2}_{-0.2}$ & Low & $>$153 \\
C-NGC\,7412-1 & 2.67 & 40.8$ \pm $3.4 & $<$0.20 & $<$2.08 & $<$3.8 & 6.6$^{+0.7}_{-0.6}$ & Medium & $>$175 \\
C-NGC\,7412-2 & 2.83 & 34.1$ \pm $3.5 & $<$0.25 & $<$2.63 & $<$4.8 & 5.8$^{+0.2}_{-0.2}$ & Low & $>$121 \\
\vspace{+1.00mm}
C-NGC\,7412-2 & 2.83 & 34.1$ \pm $3.5 & $<$0.25 & $<$2.63 & $<$4.8 & 6.6$^{+0.7}_{-0.6}$ & Medium & $>$138 \\
C-NGC\,7424 & 1.95 & 39.3$ \pm $1.7 & $<$0.15 & $<$1.58 & $<$2.9 & 16.5$^{+0.1}_{-0.1}$ & Low & $>$574 \\
C-NGC\,7424 & 1.95 & 39.3$ \pm $1.7 & $<$0.15 & $<$1.58 & $<$2.9 & 24.0$^{+0.8}_{-0.8}$ & Medium & $>$834 \\
C-NGC\,7424 & 1.95 & 39.3$ \pm $1.7 & $<$0.15 & $<$1.58 & $<$2.9 & 102.6$^{+14.3}_{-13.8}$ & High & $>$3564 \\
C-NGC\,7424 (JUN13) & 1.38 & 44.4$ \pm $1.7 & $<$0.09 & $<$0.99 & $<$1.8 & 102.6$^{+14.3}_{-13.8}$ & High & $>$5696 \\
\hline
\end{tabular}
\end{minipage}
\end{table*}

\section{Discussion}
\label{discussion}

\subsection{Low Absorption-line detection rate}
\label{discussion:detection_rate}

In Sections \ref{results_part2:hi_emission} and \ref{results_part2:hi_absorption} we found that, although the continuum source sightlines intersect the \mbox{H\,{\sc i}} disk in four of the six galaxies, no intervening \mbox{H\,{\sc i}} absorption-lines were detected.  
In this section we wish to establish the reason(s) for the lack of absorption-line detections in these galaxies.

First we address our assumption that the continuum sources are genuine background sources -- could any of the non-detections be because the continuum source is a foreground source?
Although redshifts are not available for these sources, analysis of the redshift distribution of similar radio source populations shows that the probability of them being located at redshifts $z < 0.04$ is extremely small ($\lesssim$1-2 per cent, see e.g. \citealt{1998AJ....115.1693C,2010A&ARv..18....1D}). 
\citet{2011ApJ...742...60D} calculate that, for their blind absorption-line survey, the fraction of radio sources inside their search volume ($z < 0.06$) is only 0.8 per cent. 
The flux limits and redshift space for the \citet{2011ApJ...742...60D} sample are very similar to our own survey, and we can thus we conclude that in our sample of six the chance of any of the radio sources being in the foreground is negligible.

Given this, for the sightlines intersecting the \mbox{H\,{\sc i}} disk, the absorption-line non-detection must then be due to one of the following reasons: the background continuum source being too faint, the spin temperature of the gas being too high, the covering factor being too low --- or some combination of these effects.
Since we are not able to separate the contributions of the spin temperature and covering factor with this data, we consider either the continuum flux or the \emph{ratio} of the spin temperature and covering factor, $T_{\mathrm{S}}/f$, as the two possible explanations for the non-detections.

The $T_{\mathrm{S}}/f$ limits calculated in Section \ref{results_part2:spin_temp} show that the spin temperature is likely in the normal range for spiral galaxies \citep[$\sim$100 K,][]{1995ASPC...80..357D}, in three of the four galaxies. 
Therefore, for these galaxies, the non-detections must be because the background sources are too faint to detect absorption in the current integration times. 
In the fourth case, NGC\,7424, the non-detection appears to be because $T_{\mathrm{S}}/f$ is much higher than normal, as discussed in Section \ref{results_part2:spin_temp}.

The fact that most of our non-detections are due to the continuum source being too faint is consistent with our observations of the radio sources at different resolutions. 
These show that many of the background sources are extended, or resolved into multiple components, which would have significantly reduced our absorption-line sensitivity.

\subsection{Effect of background source structure on detection-rate}
\label{discussion:background_source_structure}

\begin{figure*}
\includegraphics[width=0.30\linewidth]{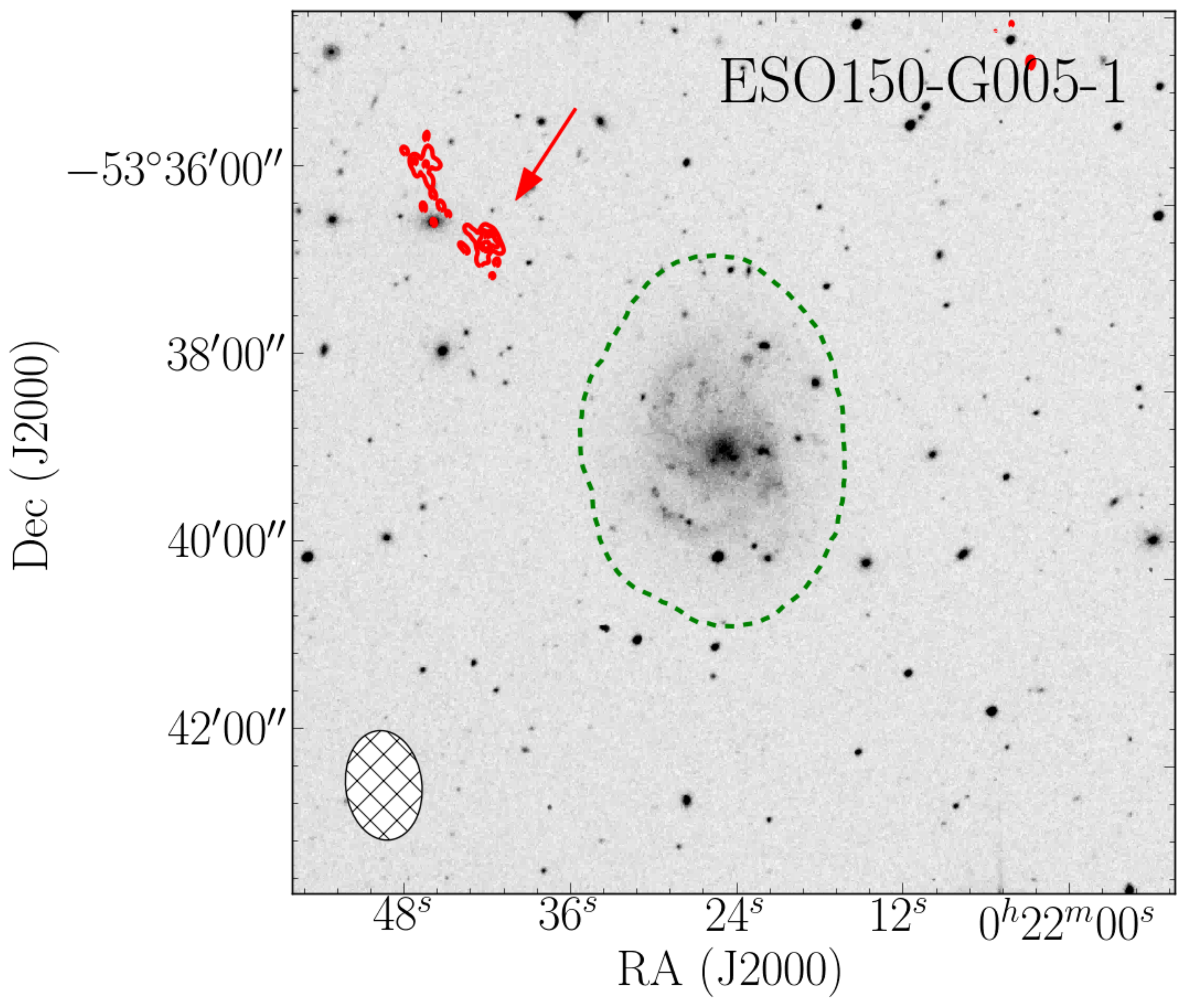}
\includegraphics[width=0.30\linewidth]{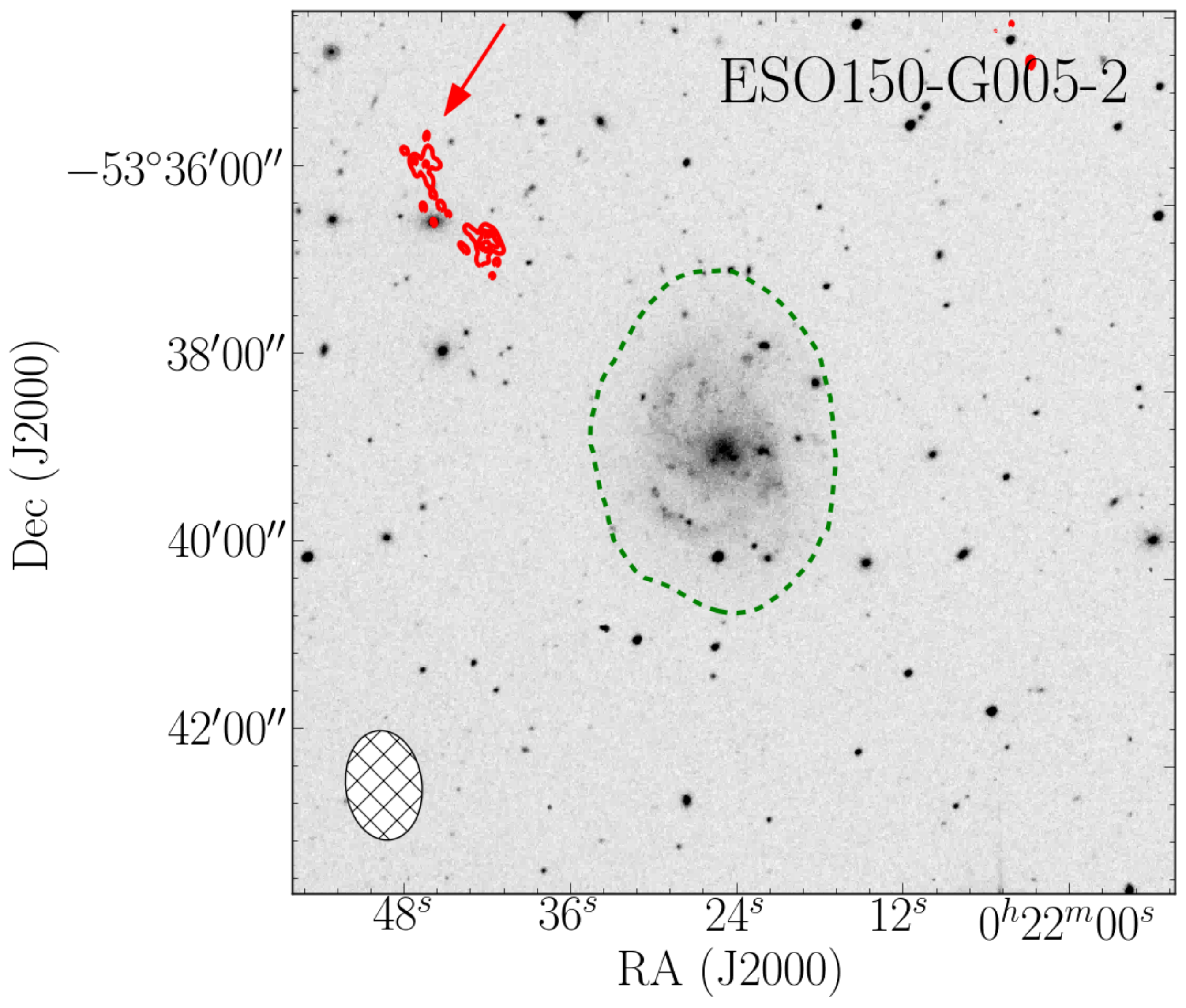}
\includegraphics[width=0.30\linewidth]{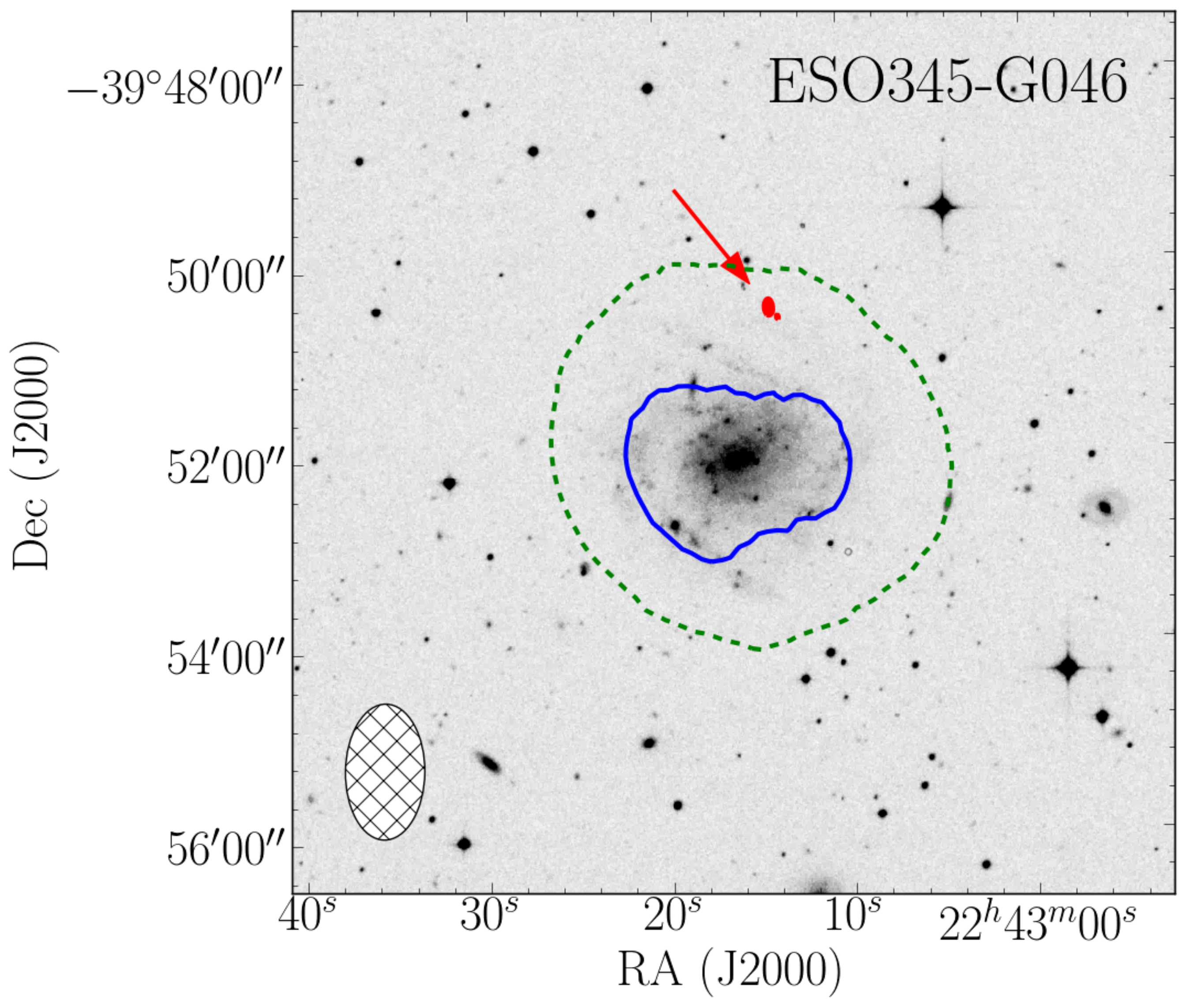}
\includegraphics[width=0.30\linewidth]{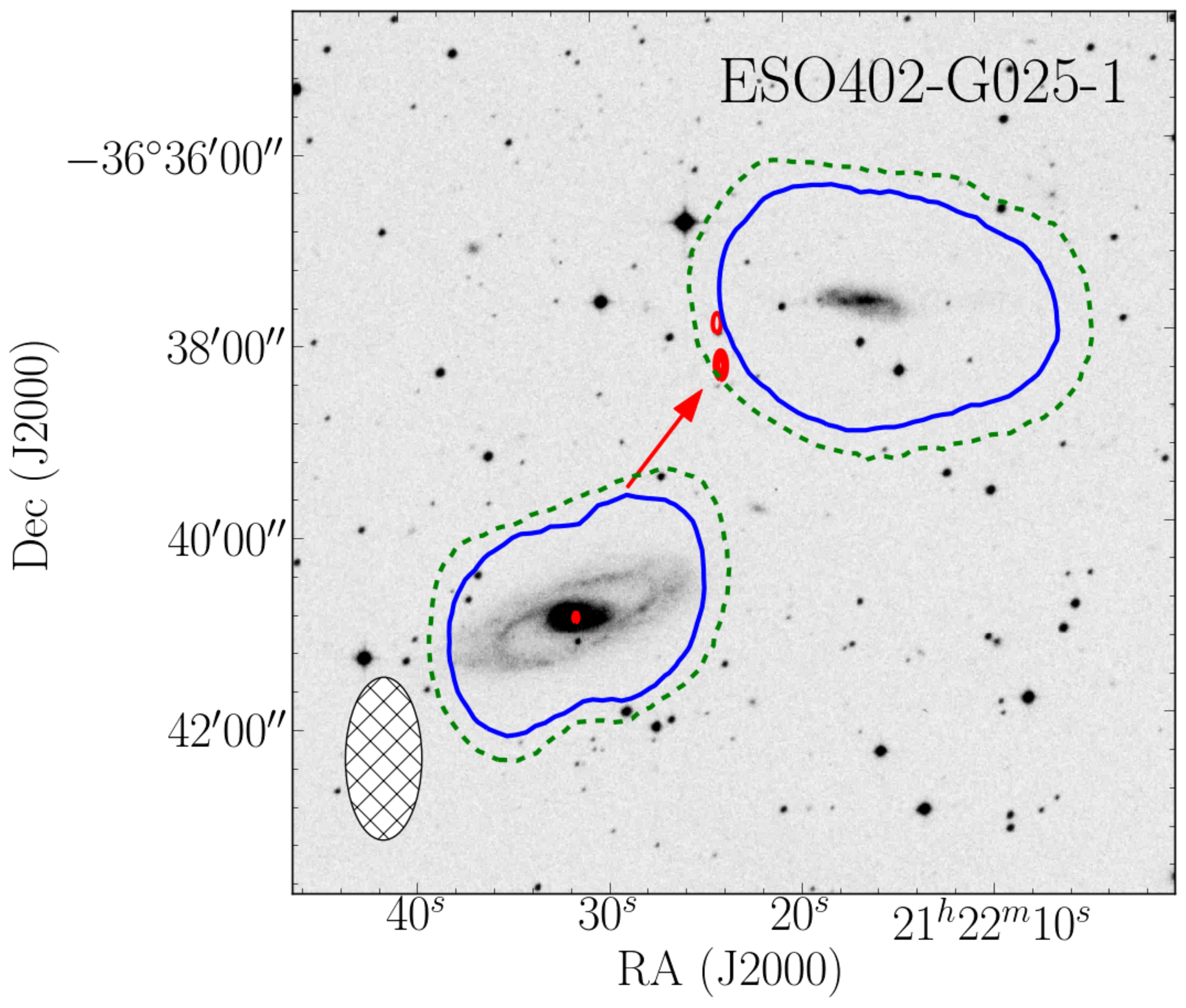}
\includegraphics[width=0.30\linewidth]{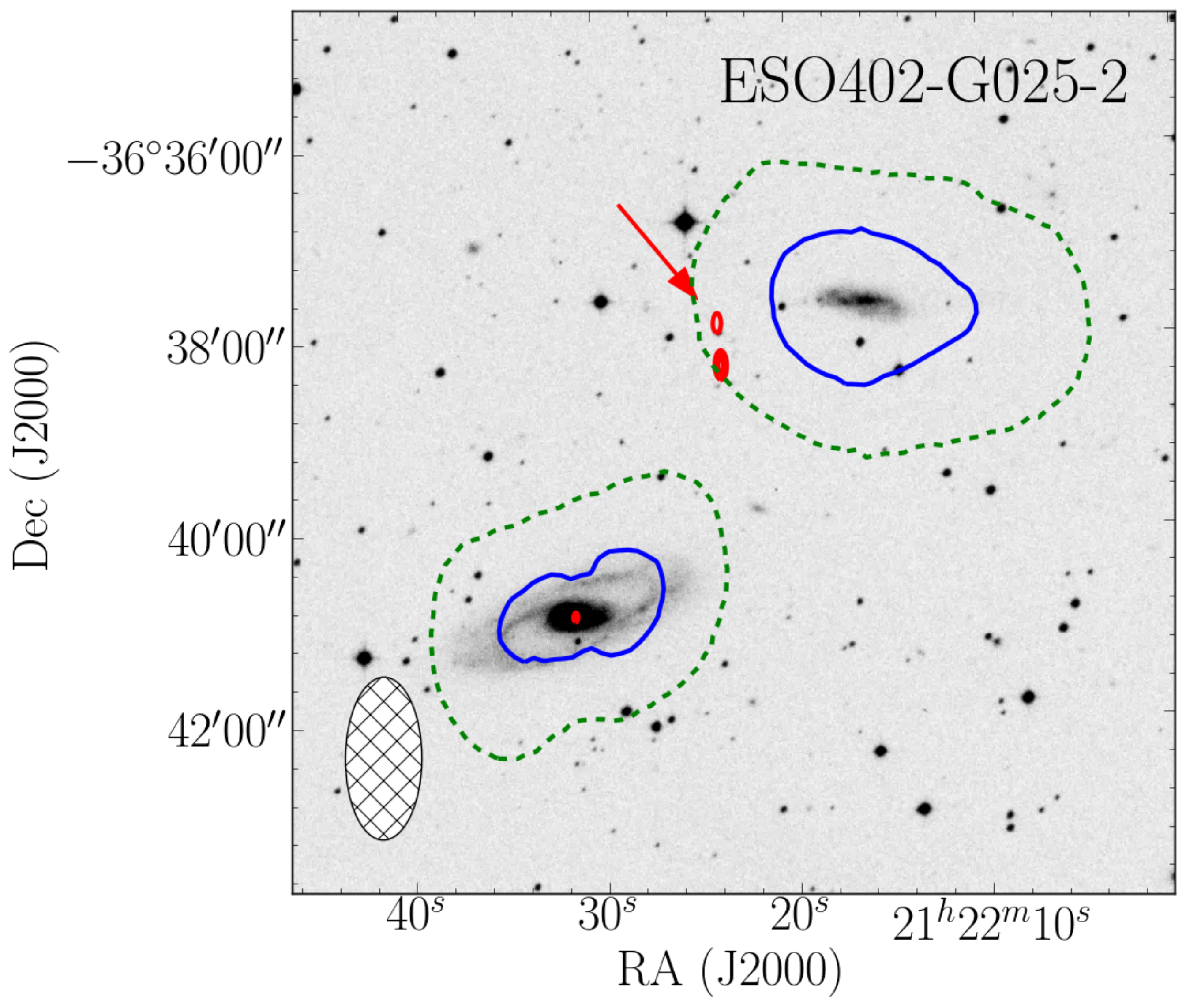}
\includegraphics[width=0.30\linewidth]{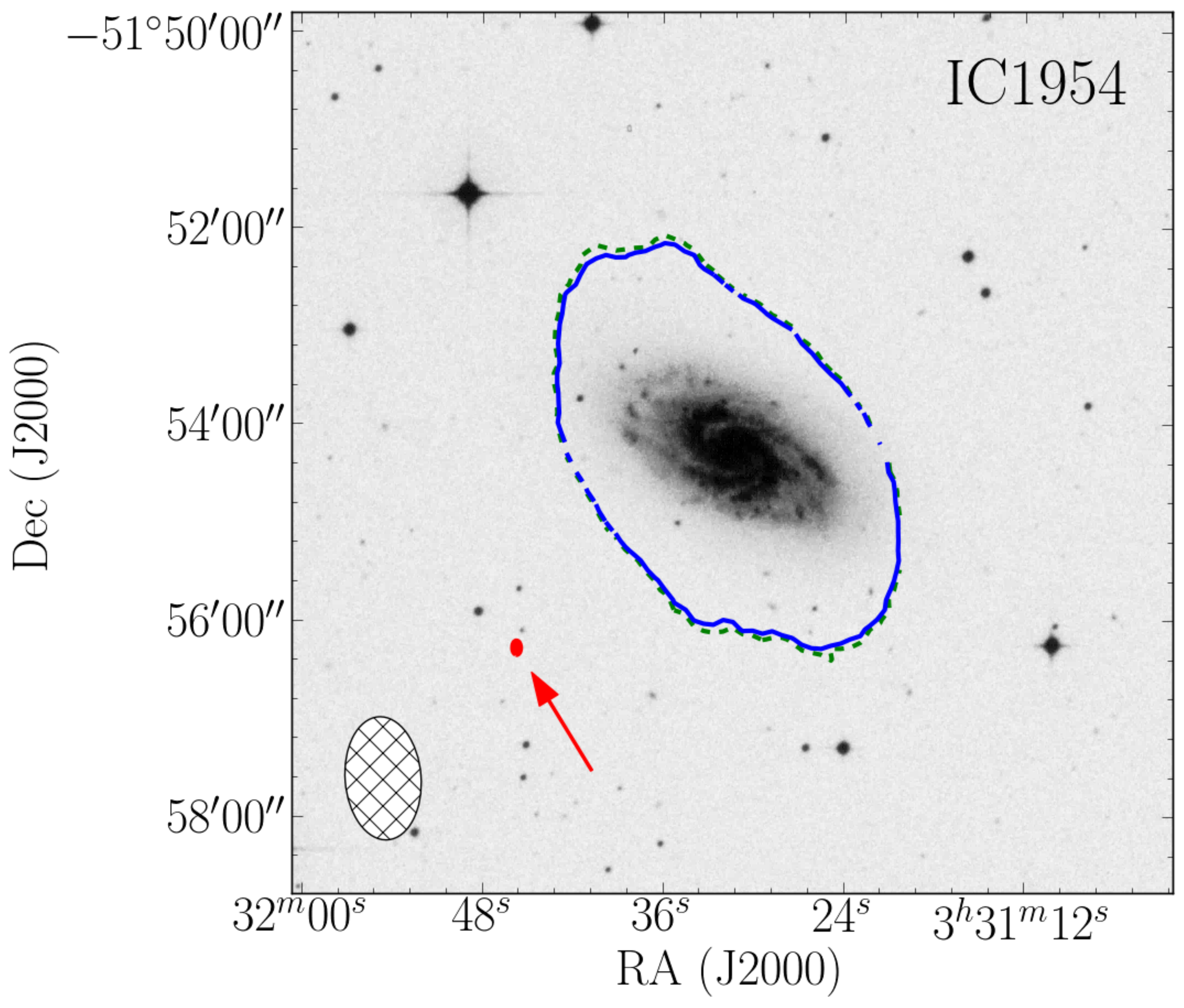}
\includegraphics[width=0.30\linewidth]{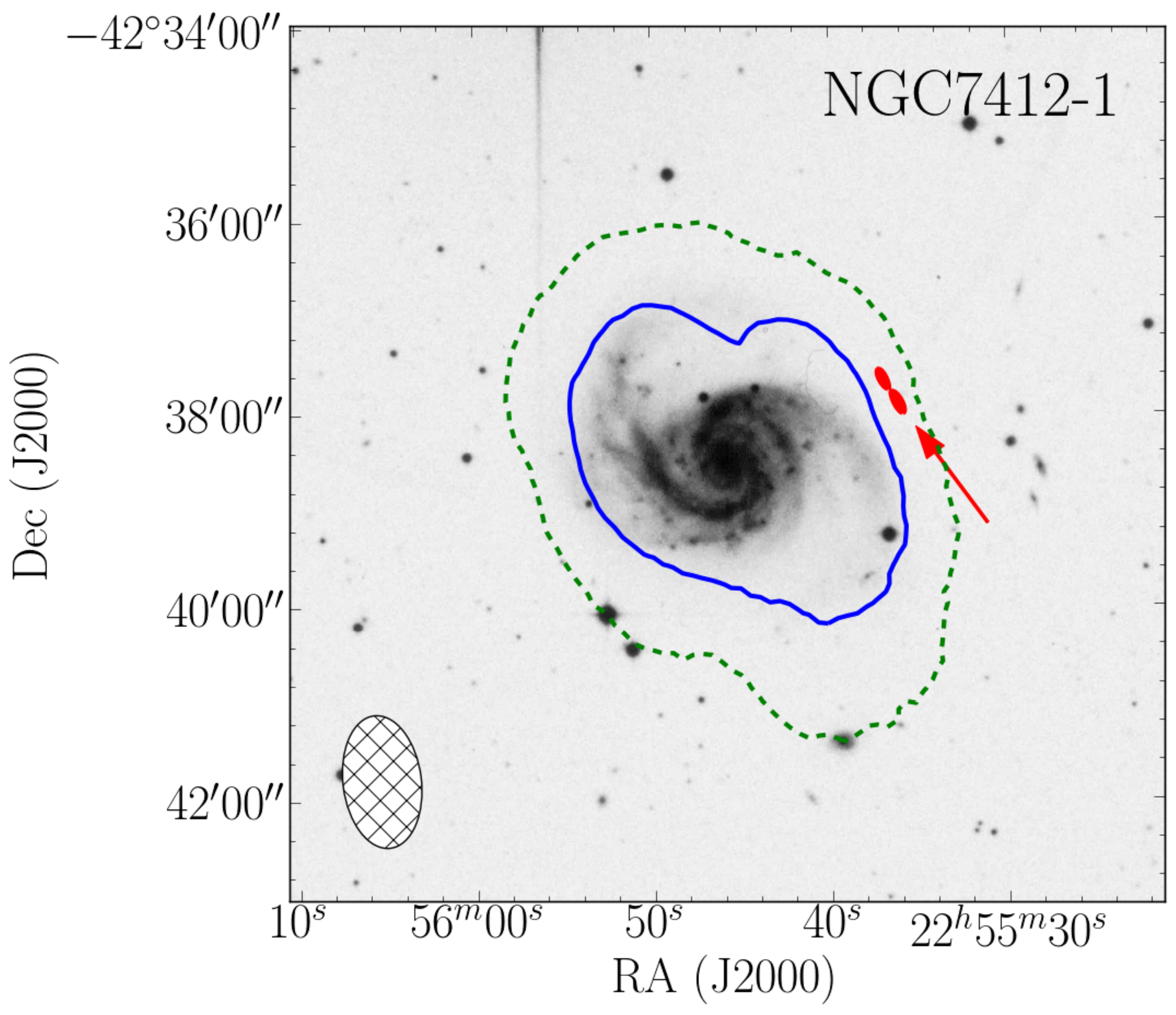}
\includegraphics[width=0.30\linewidth]{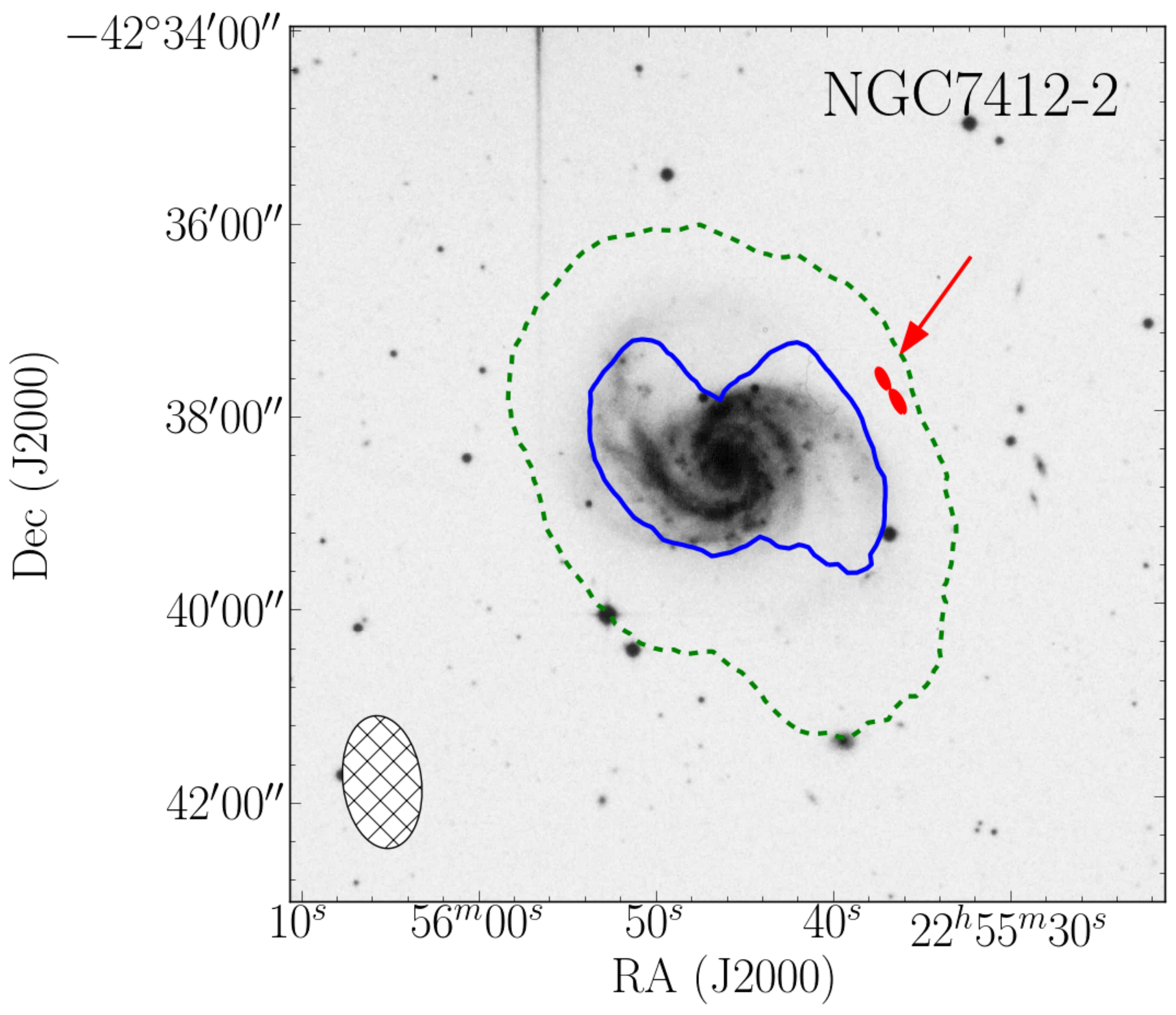}
\includegraphics[width=0.30\linewidth]{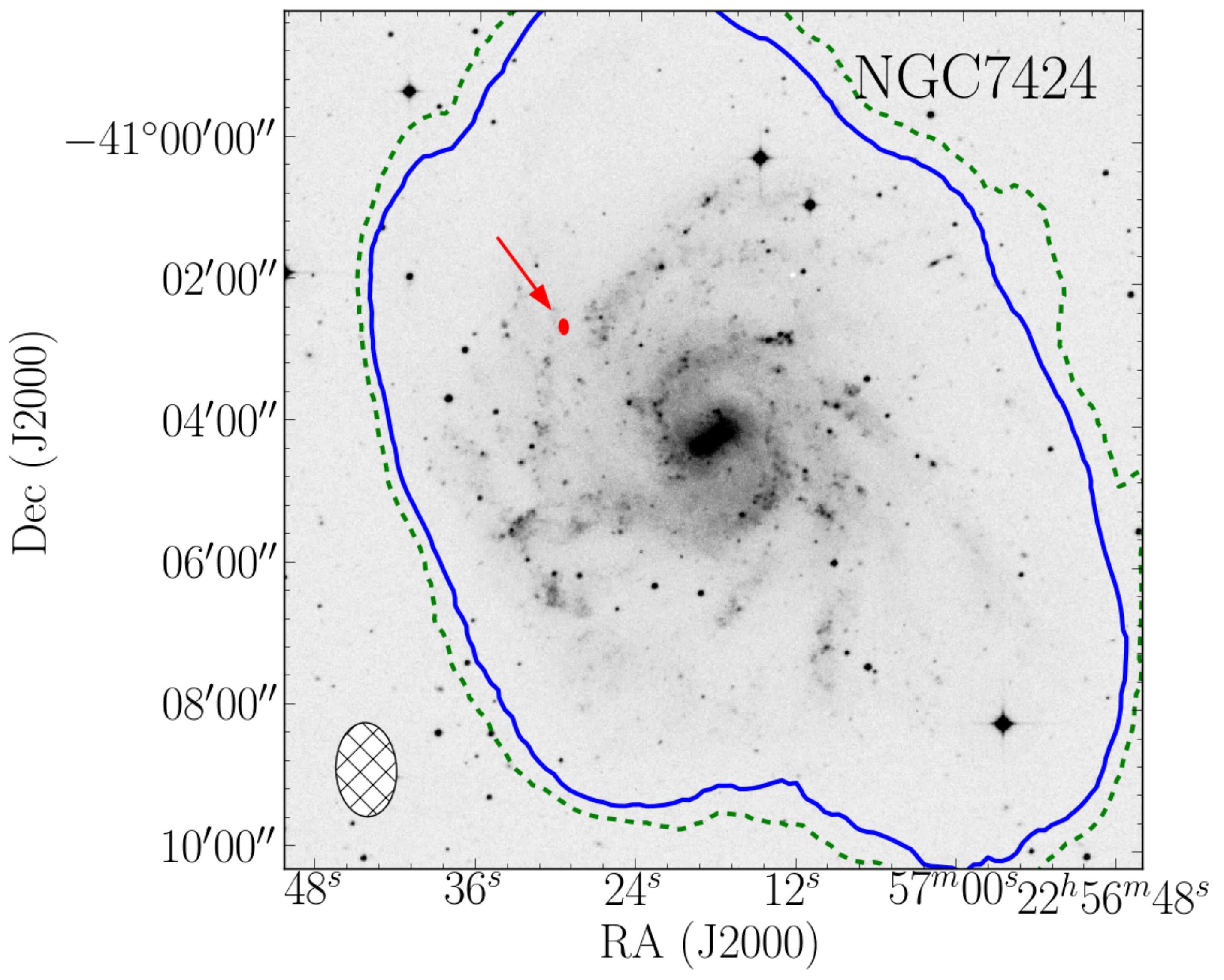}
\caption{Plots showing the absorption-line detectable region for each of the galaxies in our sample. 
The blue (solid) contour is the `actual' detectable region, given the flux measured in the high resolution image. 
The green (dotted) contour is the `expected' detectable region based on the SUMSS 843 MHz flux (assuming $\alpha$ = $-$0.7). 
For sources resolved into multiple components we have produced a separate plot for each component. 
The 1.4 GHz continuum map is also overlaid (red contours), and the arrow indicates the position of the background continuum source. 
If the continuum source lies inside the detectable region, we would have expected to detect an absorption-line (for normal values of $T_{\mathrm{S}}/f$), but if the continuum source lies outside the detectable region, it means the \mbox{H\,{\sc i}} distribution is consistent with an absorption-line non-detection.
The synthesised beam for the \mbox{H\,{\sc i}} maps is shown in the bottom left corner of each image.} 
\label{figure:column_density_contour_plots}
\end{figure*}

To further investigate the effect that the continuum source structure has had on the absorption-line sensitivity in our sample we have produced the maps shown in Figure \ref{figure:column_density_contour_plots}. 
The figures are intended to show the `absorption-line detectable region' of each of the galaxies given the flux of the background source. 
This region is determined by calculating the minimum column density detectable in absorption, assuming a 3-$\sigma$ detection, and $T_{\mathrm{S}}/f$ = 100 K. 
This value is then converted to an equivalent total \mbox{H\,{\sc i}} emission-line flux, and a contour plotted at this level on the total intensity map.

Thus, if the background source falls \emph{inside} the `detectable region' it indicates that the gas should have been detectable in absorption against that source (for the assumed $T_{\mathrm{S}}/f$). 
However, if the background source falls outside the `detectable region' then the emission-line data is consistent with the absorption-line non-detection.
The fact that C-NGC\,7424 falls inside the detectable region, despite the fact that no absorption was detected, is simply due to the higher-than-normal spin temperature in this galaxy, as discussed in Section \ref{results_part2:spin_temp}.

Two regions are plotted -- one calculated using the measured 1.4 GHz flux (in the high resolution image) and the other with the 1.4 GHz flux we would have expected from SUMSS (assuming $\alpha$ = $-$0.7) if the source was unresolved. 
In almost all cases the `actual' detectable region (in blue) is significantly smaller than the `expected' detectable region (in green), and in some cases has disappeared entirely (if there is no blue contour, this indicates that the column density is not high enough anywhere in the galaxy to be detectable in absorption, against that particular source).

This shows that the loss of continuum flux has severely affected the ability to detect absorption, especially where the continuum source is located further out in the disk. 
Specifically, we find that in several cases we would have expected to detect absorption in several of the galaxies, based on their SUMSS fluxes, but given their actual fluxes we are in fact significantly below the detection limit.
We note also that if the continuum sources have structure on scales still not resolved in our highest resolution ATCA images, then the `actual' (blue) detectable region will be even smaller than shown. 
Table \ref{table:results_summary} gives a summary of these results.

We now consider what effect this might have on future large absorption-line surveys, such as FLASH. 
FLASH will survey the entire southern sky, searching for \mbox{H\,{\sc i}} absorption against 150,000 radio continuum sources (all radio sources brighter than about 50 mJy at 1.4 GHz). 
The FLASH sample will therefore contain an abundance of very bright (Jy-level) radio sources, ideal for detecting \mbox{H\,{\sc i}} absorption. 
However, as we have found in this work, the continuum sources may have structure not seen in the lower resolution surveys from which they will be selected, which would reduce the peak flux and affect the absorption-line sensitivity.

For the brightest sources this will not matter provided the source flux remains above the absorption-line detection limit. 
Similarly, sources already too faint to detect absorption against will be unaffected. 
However, this effect will matter in systems which are on the borderline of detectability -- where an unresolved source might be bright enough to detect absorption against, but if the source turns out to be resolved it could cause the system to drop below the detection limit.

Our sample size is small, so further work needs to be done with much larger sample sizes. 
However if, as we have found in this sample, a large fraction of the radio source population are extended on the scales at which \mbox{H\,{\sc i}} absorption occurs, this could affect the detection rate in future surveys. 
Given the relatively low spatial resolution of existing all-sky continuum surveys, we should also investigate other methods of gaining information about the source structure -- for example using spectral index as a proxy for compactness (see \citealt{2011MNRAS.412..318M}). 

\begin{table*}
\begin{minipage}{\linewidth}
\centering
\caption{Summary of the emission- and absorption-line results. 
Column (1) is the sightline searched. 
Columns (2) and (3) state whether an emission-line or absorption-line, respectively, were detected along this sightline. 
Column (4) gives the likely reason that absorption was not detected along that sightline. 
Column (5) states whether we would have expected to detect \mbox{H\,{\sc i}} absorption given the SUMSS flux and the observed \mbox{H\,{\sc i}} distribution. 
Column (6) gives some properties of the background source structure.}
\label{table:results_summary}
\begin{threeparttable}
\begin{tabular}{@{} llllll @{}} 
\hline
Sightline & Emission- & Absorption- & Reason for & Detection & Radio source\\
& line? & line? & non-detection\tnote{$*$} & expected (SUMSS)? & \\
\hline
C-ESO\,150-G\,005-1 & No & No & ND & No & Extended \\
C-ESO\,150-G\,005-2 & No & No & ND & No & Extended \\
C-ESO\,345-G\,046 & Yes & No & CF & Yes & Single component \\
C-ESO\,402-G\,025-1 & Yes & No & CF & Borderline & Double \\
C-ESO\,402-G\,025-2 & Yes & No & CF & Borderline & Double \\
C-IC\,1954 & No & No & ND & No & Single component \\
C-NGC\,7412-1 & Yes & No & CF & Yes & Double \\
C-NGC\,7412-2 & Yes & No & CF & Yes & Double \\
C-NGC\,7424 & Yes & No & $T_{\mathrm{S}}/f$ & Yes\tnote{$\dagger$} & QSO, point source \\
\hline
\end{tabular}
\begin{tablenotes}
\footnotesize{
\item[$*$] {Reason for non-detection:}
\item[] {`ND' means that the sightline does not intersect the disk.}
\item[] { `CF' means that the continuum flux of the background source was too low. The emission- and absorption-line data are consistent with each other for normal values $T_{\mathrm{S}}/f$.}
\item[] {`$T_{\mathrm{S}}/f$' means that the ratio of $T_{\mathrm{S}}/f$ must be much higher than normal to explain the absorption-line non-detection (given the observed \mbox{H\,{\sc i}} distribution).}
\item[$\dagger$] {This result assumes $T_{\mathrm{S}}/f$ = 100 K. However, since $T_{\mathrm{S}}/f$ is likely considerably higher than 100 K for NGC\,7424, in reality, the continuum sightline probably would not intersect the expected detectable region.}
}
\end{tablenotes}
\end{threeparttable}
\end{minipage}
\end{table*}

\subsection{Detection rate as a function of impact parameter}
\label{discussion:detection_rate_impact_parameter}

\begin{figure}
\includegraphics[width=0.475\textwidth]{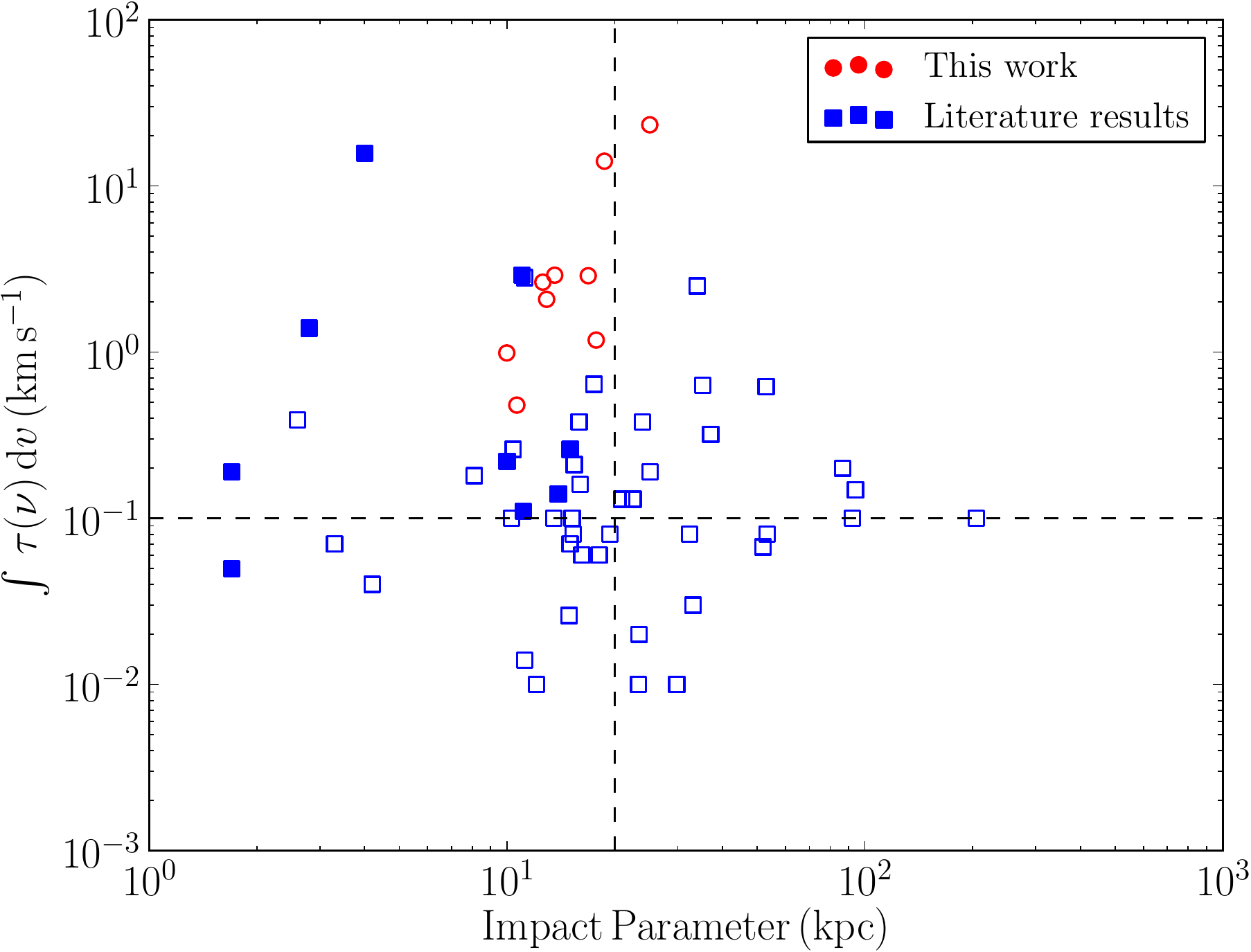}
\caption{Plot of integrated optical depth vs. impact parameter for different intervening absorption-line searches. 
The results of our survey are indicated by circles (red), and the literature results (quasar sightlines only) by squares (blue). 
Detections are shown as filled markers, and non-detections as open markers. 
Non-detections are plotted as the 3-$\sigma$ upper limit on the integrated optical depth.}
\label{figure:gupta_comparison}
\end{figure}

In Figure \ref{figure:gupta_comparison} we plot the integrated optical depth against the impact parameter for each of the sightlines in our sample. 
For comparison, we also show results from previous intervening absorption-line surveys. 
The results shown comprise the list compiled by \citet{2010MNRAS.408..849G} (\citealt{1975ApJ...200L.137H,1988A&A...191..193B,1990ApJ...356...14C,1992ApJ...399..373C,2004ApJ...600...52H,2010ApJ...713..131B}), as well as other more recent results \citep{2013MNRAS.428.2198S,2011ApJ...727...52B,2014ApJ...795...98B}. 
We note that, for surveys which include both radio galaxies and quasars, only the quasar sightlines are shown (as per \citealt{2010MNRAS.408..849G}) -- a point that we discuss in further detail below.

\citet{2010MNRAS.408..849G} estimated that there is a 50 per cent chance of making an absorption-line detection at impact parameters of 20 kpc or less, for integrated optical depths greater than 0.1 km s$^{-1}$ (i.e. the upper left quadrant of Figure \ref{figure:gupta_comparison}). 
More recent results show similar detection rates. 
However, although all but one of our sightlines fall in this quadrant, we have not made any detections (we note though that very narrow components such as that detected by \citet{2014ApJ...795...98B} would not be detectable in our survey due to the much lower spectral resolution.)

One likely explanation for the apparent discrepancy between the results of ours and previous surveys lies in the type(s) of continuum sources used to search for absorption. 
In our sample only one of the background continuum sources is a known quasar, while the remaining sources are unclassified and therefore most likely to be radio galaxies (a fact that is further supported by the double structure seen in a number of cases). 
In contrast, all of the literature results in Figure \ref{figure:gupta_comparison} are from quasar sightlines (which provide very bright radio sources, ideal for targeted \mbox{H\,{\sc i}} absorption searches).

Quasars are far more likely to be compact radio sources, compared with radio galaxies (hence why they are frequently used for targeted absorption searches), and we might therefore expect that the detection rate along quasar sightlines would be higher.
Since, in the overall radio source population, only about 10 per cent of sources are quasars, our results may be more representative of the expected detection rate for blind absorption-line surveys, such as FLASH, in the future. 
The lack of detections in our sample at impact parameters $<$20 kpc therefore raises the question of whether the true detection rate expected for blind, unbiased samples would be as high as 50 per cent. 
Expanding our sample size would allow us to investigate this effect further.

Although different surveys show different detection rates for impact parameters $<$20 kpc, it seems clear that the detection rate at impact parameters greater than 20 kpc is likely to be very low. 
The radial profiles presented in Section \ref{results_part1:radial_profiles} can provide further information about this. 
For our galaxies, we find that the \mbox{H\,{\sc i}} column density typically falls below the DLA limit of 2 $\times$ 10$^{20}$ at a radius of around 10-20 kpc (see Table \ref{table:HI_opt_disk_sizes}). 
Since these are the kind of gas-rich systems that FLASH, and other blind surveys, are most likely to detect, this again suggests that most of the detections made will be along sightlines with impact parameters of 20 kpc or less.

\section{Conclusions}
\label{conclusions}

We have conducted a targeted search for intervening \mbox{H\,{\sc i}} absorption in six nearby, gas-rich galaxies, using observations with the ATCA. 
By targeting nearby galaxies we obtained both \mbox{H\,{\sc i}} emission- and \mbox{H\,{\sc i}} absorption-line data, allowing us to directly relate the neutral gas distribution to the absorption-line detection rate.

We have produced the first resolved \mbox{H\,{\sc i}} maps of the target galaxies. 
We find that four of the sightlines searched intersect the \mbox{H\,{\sc i}} disk at column densities above 1-2 $\times$ 10$^{20}$ cm$^{-2}$, however, no absorption was detected. 
From analysis of the \mbox{H\,{\sc i}} emission-line maps, we conclude that the non-detections for three of these four galaxies are simply because the background sources are too faint to detect absorption in the current integration times.

For the fourth galaxy, NGC\,7424 we obtain a limit of $T_{\mathrm{S}}/f$ $>$ 5700 K, which explains why we have not detected any absorption, despite the high \mbox{H\,{\sc i}} column density along this sightline. 
We suggest that the actual spin temperature may not be significantly higher than normal if the covering factor is low due to a clumpy \mbox{H\,{\sc i}} medium. 
Limits obtained for the remaining galaxies in our sample typically yield estimates of $\sim$100-200 K, consistent with the expectation for normal spiral galaxies, and with other recent results.

Many of the background sources in our sample became extended, or resolved into multiple components in the high resolution ATCA images, reducing our absorption-line sensitivity. 
Plots of the `absorption-line detectable region' illustrate how dramatically this can affect the probability of detecting absorption, particularly in the outer parts of the disk, and also highlight the effect that background source structure might have on the detection rate of future large absorption-line surveys. 
Given the relatively low spatial resolution of existing all-sky radio surveys, other methods for estimating the compactness of continuum sources, such as spectral index properties, should be investigated, in preparation for future surveys.

Our detection rate is low compared with that of previous surveys, and we find that this may be explained by differences in the type(s) of radio sources targeted by different surveys. 
Samples targeting quasar sightlines, which form the majority of surveys to date, would be expected to have a higher detection rate, as quasars are normally very compact, maximising the absorption-line sensitivity. 
Given that our sample consists mainly of radio galaxies, which as we have seen above are often more extended, it is perhaps not surprising that our detection rate is lower. 
Since quasars represent only about 10 per cent of the overall radio source population, we suggest that our results may therefore be more representative of the expected detection rate for large, blind absorption-line surveys in the future. 
Given our small sample size, however, further observations would be required to confirm whether this is the case.

Radial \mbox{H\,{\sc i}} profiles of the target galaxies provide additional information about the extent of the \mbox{H\,{\sc i}} disks. 
We find that the column density typically drops below the DLA-limit (2 $\times$ 10$^{20}$ cm$^{-2}$) at a radius of 10-20 kpc, suggesting that the majority of detections in future blind surveys will be at impact parameters of 20 kpc or less. 
This is consistent with absorption-line results to date, which show detections at impact parameters greater than 20 kpc to be extremely rare.
Future work including an expanded sample, and VLBI observations would allow us to better quantify the expected detection rate, and the effect of background source structure for the next generation of absorption-line surveys.

\section{Acknowledgements}
\label{acknowledgements}

SNR acknowledges financial support from an Australian Postgraduate Award. 
JRA acknowledges financial support from an ARC Super Science Fellowship and a CSIRO Bolton Fellowship. 
The authors wish to thank Bjorn Emonts for useful discussions and assistance preparing for the observations. 
We also thank Tom Oosterloo, Raffaella Morganti and other staff at ASTRON for useful discussions and suggestions. 
The Australia Telescope and Parkes radio telescope are funded by the Commonwealth of Australia for operation as a 
National Facility managed by CSIRO.
The Centre for All-sky Astrophysics is an Australian Research Council Centre of Excellence, 
funded by grant CE110001020. 
This research has made use of the NASA/IPAC Extragalactic Database (NED) which is operated by the Jet Propulsion Laboratory, California Institute of Technology, under contract with the National Aeronautics and Space Administration, and the SIMBAD database, operated at CDS, Strasbourg, France. 
This research made use of APLpy, an open-source plotting package for Python hosted at http://aplpy.github.com, and Astropy, a community-developed core Python package for Astronomy (Astropy Collaboration, 2013).
The authors have also made use of NASAs Astrophysics Data System Bibliographic Services.

\appendix

\section[]{H\,{\sevensize\bf I} moment maps}
\label{appendix:moment_maps}

This appendix presents the \mbox{H\,{\sc i}} moments maps for the galaxies in our sample. 
For each galaxy we show the zeroth, first, and second moment maps (\mbox{H\,{\sc i}} total intensity, mean velocity, and velocity dispersion maps, respectively). 
The \mbox{H\,{\sc i}} total intensity maps are also shown in Figure \ref{figure:overlay_maps}, overlaid as contours on the optical image, and were used in producing the radial \mbox{H\,{\sc i}} profiles in Figure \ref{figure:radial_profiles} and the absorption-line detectable regions in Figure \ref{figure:column_density_contour_plots}.

\begin{figure*}
\includegraphics[width=\linewidth]{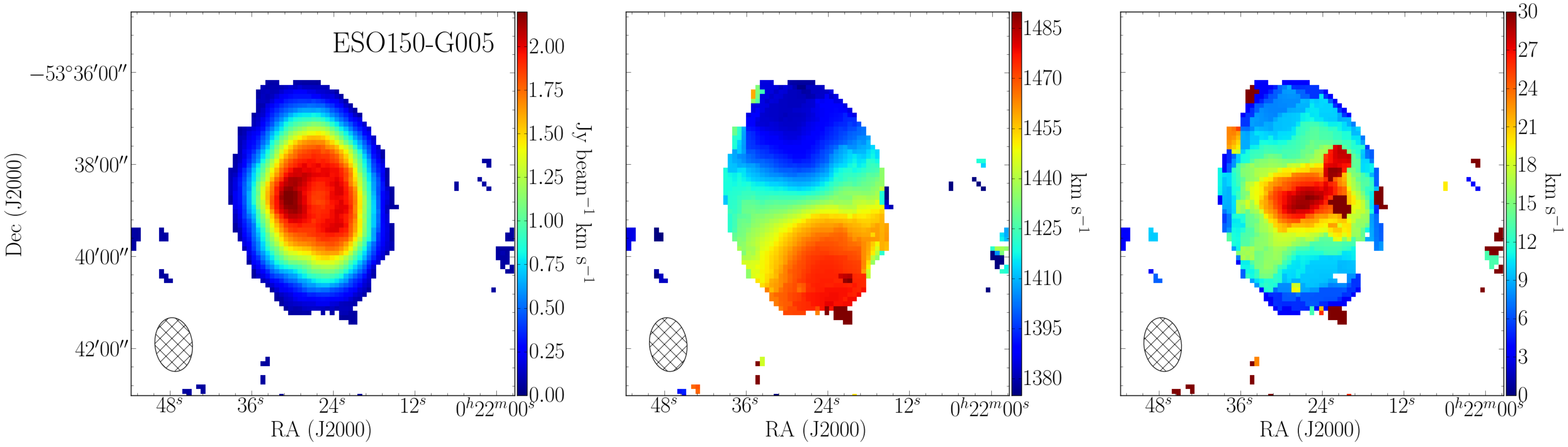}
\includegraphics[width=\linewidth]{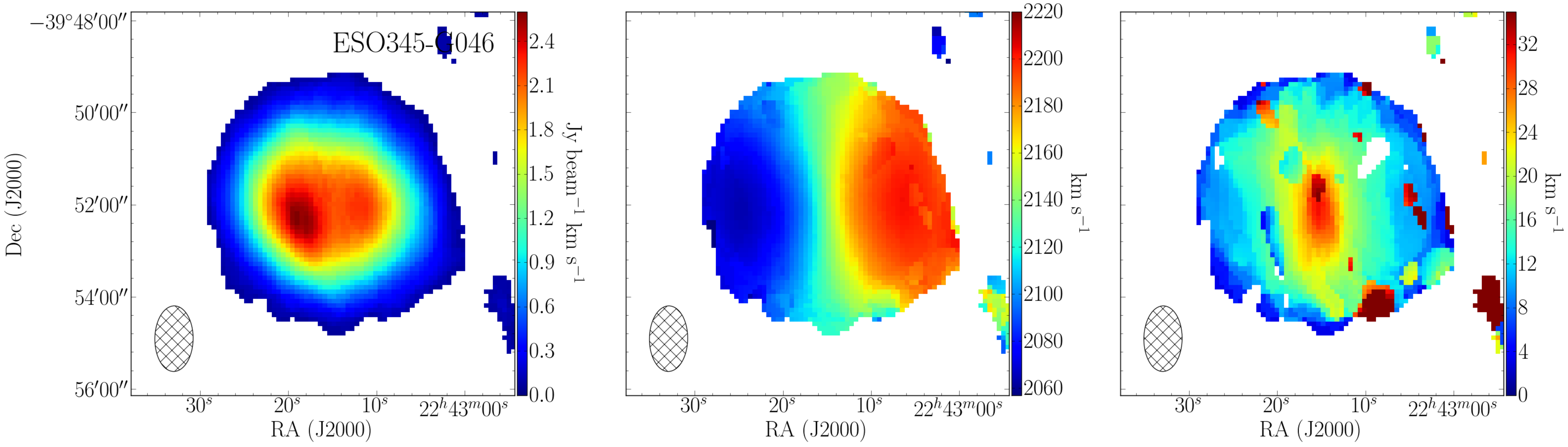}
\includegraphics[width=\linewidth]{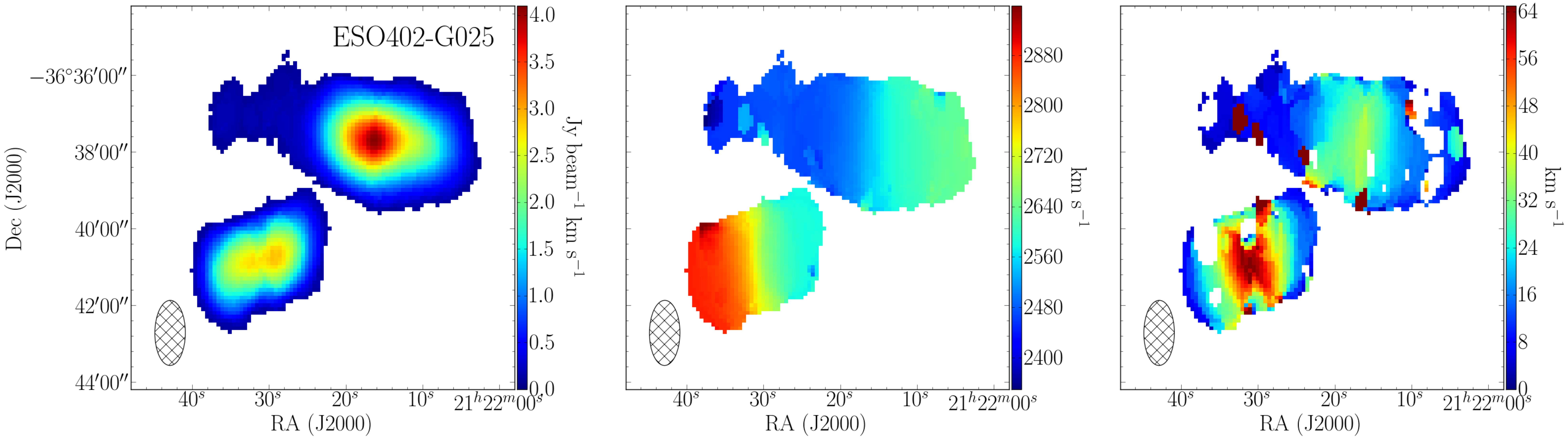}
\includegraphics[width=\linewidth]{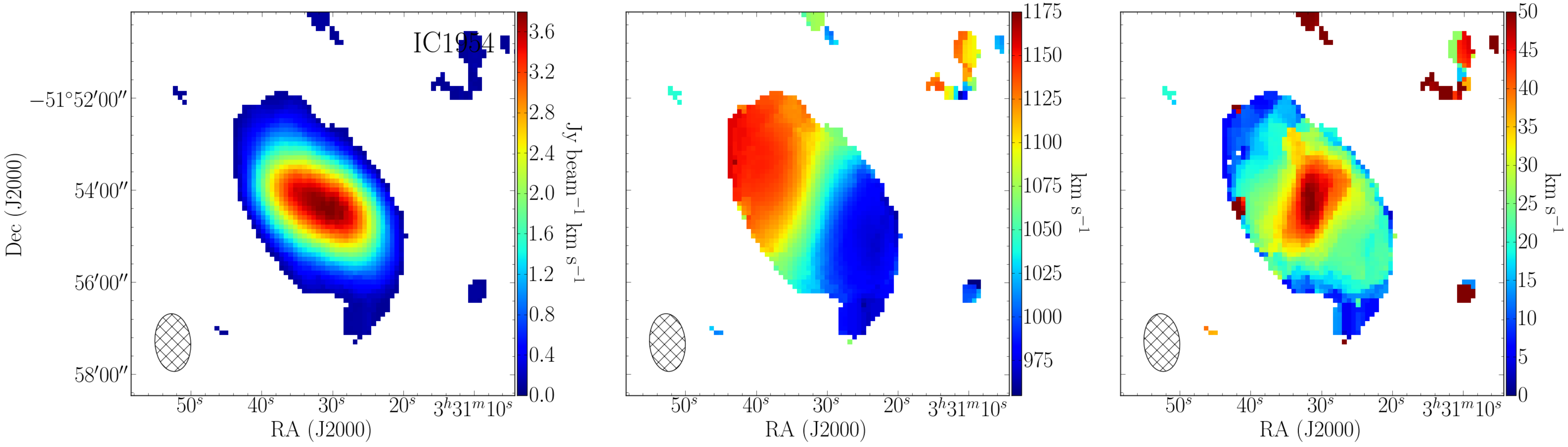}
\caption[]{\mbox{H\,{\sc i}} moment maps of the target galaxies (produced from the low resolution cubes). 
Left to right: \mbox{H\,{\sc i}} total intensity (zeroth moment), mean \mbox{H\,{\sc i}} velocity (first moment), and \mbox{H\,{\sc i}} velocity dispersion (second moment) maps.
The synthesised beam is shown in the bottom left corner of each image.}
\label{figure:moment_maps}
\end{figure*}

\begin{figure*}
\includegraphics[width=\linewidth]{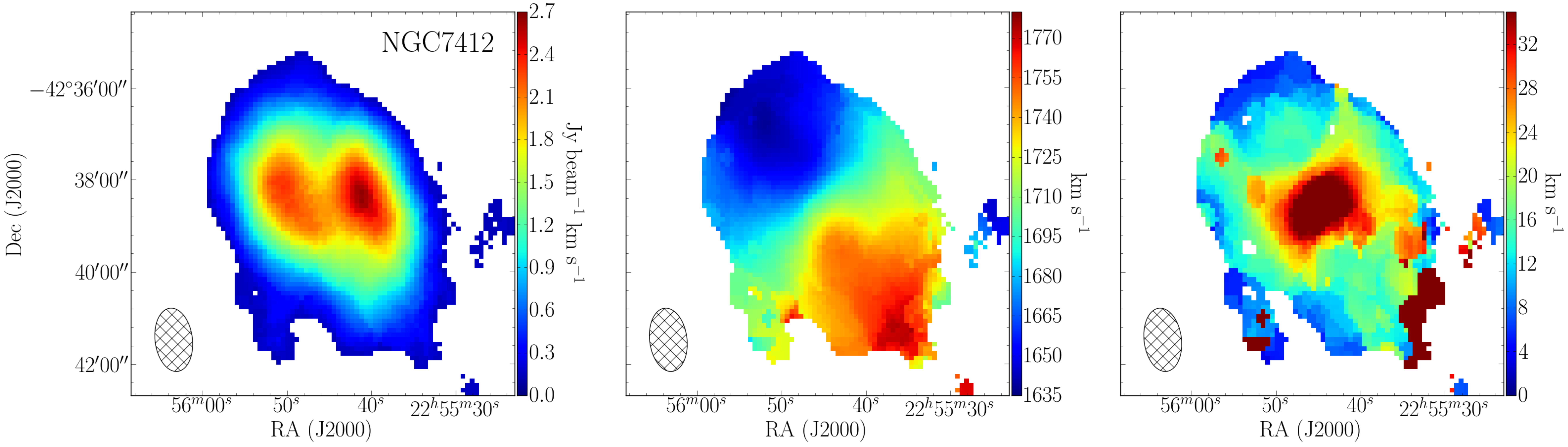}
\includegraphics[width=\linewidth]{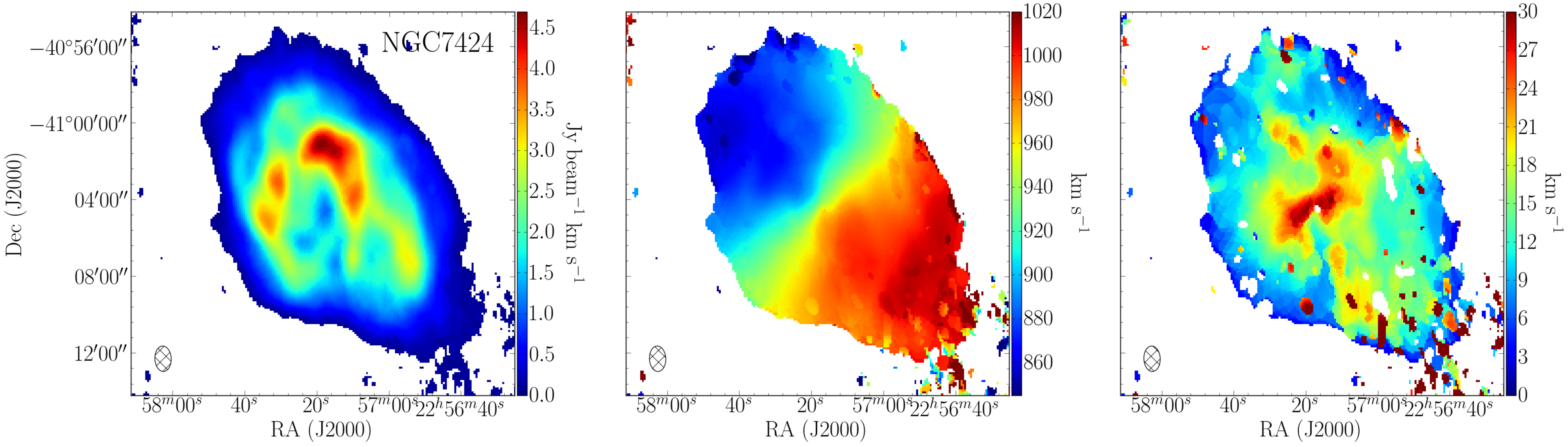}
\contcaption{}
\end{figure*}

\section[]{SN2001ig in NGC\,7424}
\label{appendix:sn2001ig}

\begin{figure*}
\includegraphics[width=\textwidth]{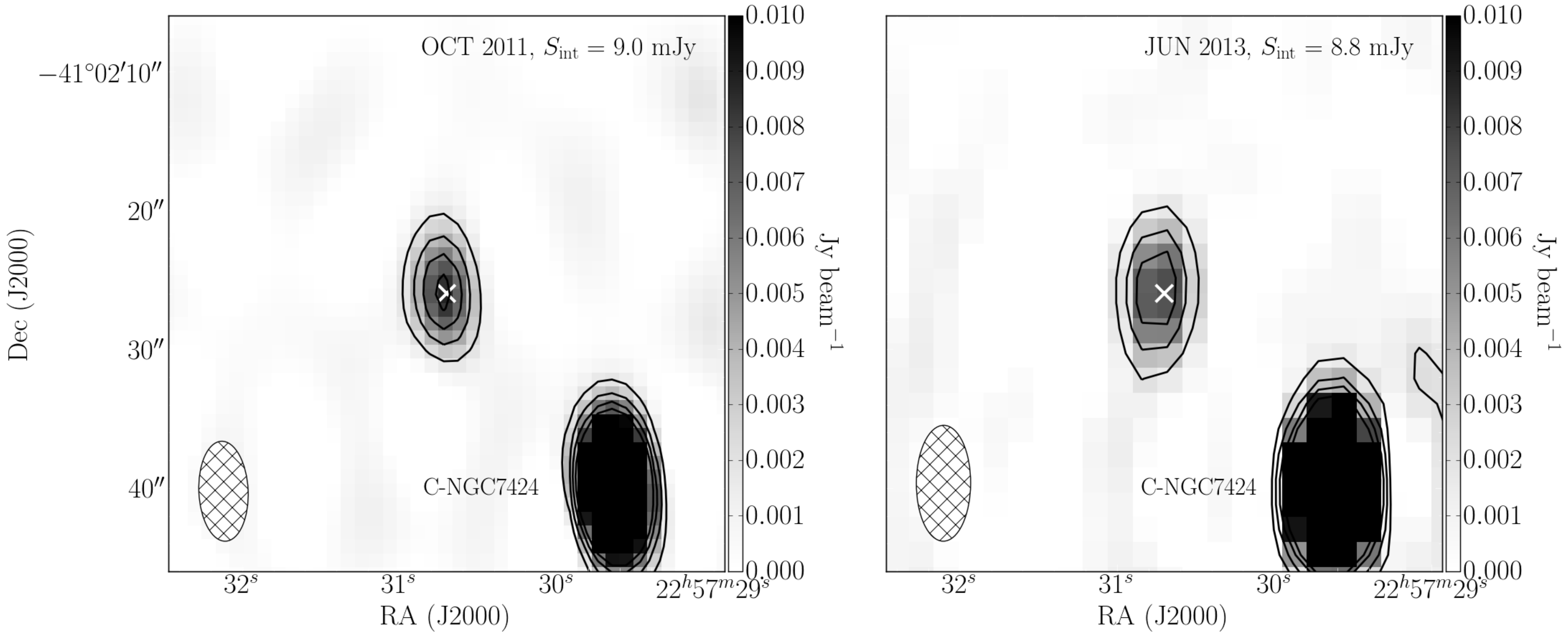}
\caption{SN2001ig in NGC\,7424 (1.4 GHz continuum images). 
\emph{Left:} October 2011 image. \emph{Right:} June 2013 image. 
The optical position of the supernova is indicated by the white cross. 
Both images are plotted over the same region of sky and with the same colour-scale, to allow easy comparison. 
The total flux at each epoch is given in the top right corner and the contour levels are set at 2.0, 4.0, 6.0, and 8.0 mJy beam$^{-1}$. 
The synthesised beam is indicated by the ellipse in the bottom left corner. 
The other radio source in the image is C-NGC\,7424, the quasar used to search for intervening \mbox{H\,{\sc i}} absorption in the disk of NGC\,7424.}
\label{figure:sn2001ig}
\end{figure*}

NGC\,7424 is the host of the supernova SN2001ig \citep[22:57:30.7 $-$41:02:26,][]{2001IAUC.7772....1E}, which is coincidentally located very close to the quasar being used to search for absorption in the disk of NGC\,7424. 
The supernova has a known radio counterpart \citep{2001IAUC.7777....2R}, and the radio light curve was studied in detail for at least 700 days after the initial supernova event \citep{2004MNRAS.349.1093R}. 

We report the re-detection of the radio counterpart, more than 10 years after the initial supernova event. 
A faint radio source was detected in both the October 2011 and June 2013 high resolution ATCA continuum images, at the optical position of the supernova. 
In both cases the source was unresolved, with 1.4 GHz fluxes of 9.0 $\pm$ 0.6 and 8.8 $\pm$ 0.7 mJy, respectively. 
The images for the two epochs are shown in Figure \ref{figure:sn2001ig}.

\bsp

\label{lastpage}

\end{document}